\documentclass[twocolumn,amsmath,amssymb]{revtex4}
\usepackage{epsfig,epsf,graphics,color,rotate}
\usepackage{dcolumn}
\usepackage{color}
\usepackage{amsmath}
\usepackage{amsfonts}
\usepackage{epsfig}
\begin{document}

\title{PIV mapping of pressure and velocity fields in the plane magnetohydrodynamic Couette flow.
}


\author{B. Moudjed$^{1,2}$, A. Poth\'erat$^1$ and M. Holdsworth$^1$}


\affiliation{$^1$Fluid and Complex Systems  Research Centre, Coventry University, Priory Street,
              Coventry CV1 5FB, United Kingdom \\
$^2$LNCMI-EMFL-CNRS, UGA, 25 Avenue des Martyrs, 38000 Grenoble, France
              \email{brahim.moudjed@yahoo.fr}
              \email{alban.potherat@coventry.ac.uk}
}

\date{\today}

\begin{abstract}
We present the first simultaneous mapping of two-dimensional, time-dependent velocity and pressure fields
 in a plane Couette flow pervaded by a transverse magnetic field. While electromagnetic forces are
strongest in fluids of high electric conductivity such as liquid metals, their opacity
excludes optical measurement methods. We circumvent this difficulty using a transparent electrolyte (Sulphuric acid), whose weaker conductivity is offset by higher magnetic fields. We describe an experimental rig
based on this idea, where the Couette flow is entrained by a tape immersed in sulphuric acid and
positioned flush onto the bore of large superconducting magnet, so that most of the flow is
pervaded by a sufficiently homogeneous transverse magnetic field. Velocity and pressure fields are obtained
by means of a bespoke PIV system, capable of recording the fluid's acceleration as well as its velocity.
Both fields are then fed into a finite difference solver that extracts the pressure field from the
magnetohydrodynamic governing equations. This method constitutes the first implementation of the pressure PIV technique to an MHD flow. Thanks to it, we obtain the first experimental velocity and pressure
profiles in an MHD Couette flows and show that the transitional regime between laminar and turbulent states is dominated by near-wall, isolated, anisotropic perturbations.
~                                                                   
\end{abstract}
\maketitle

\section{Introduction}
\label{sec:intro}
We report the first mapping of velocity and pressure fields in a plane Couette flow pervaded by a
transverse magnetic field, also called the \emph{plane MagnetoHydroDynamic (MHD) Couette flow}.
The Couette flow is one of the classical problems of fluid mechanics. It owes its popularity to its
generic simplicity - a unidirectional flow driven by the shear due to the relative movement of
its boundaries - and to its relevance to a very wide range of ideal and practical problems
(lubrication, local shear in turbulent flows, atmospheric, geophysical  phenomena etc..). Despite its apparent simplicity, the Couette flow, with or without magnetic field remains an ideal
playground to understand transition to turbulence and sheared turbulence, two of the major challenges
of fluid dynamics \cite{grossmann2016_arfm}.
Nevertheless, a good experimental approximation to the ideal Couette flow -\emph{i.e} the flow between
two infinitely extended planes, unspoilt by unwanted physical phenomena - is very difficult of achieve.
Difficulties arise because of its infinitely extended geometry,
which plays a key role in transition mechanisms, especially in the development of
patterns of localised turbulence \cite{duguet2010_jfm,barkley2005_prl}.
Only recently did experimental work show that the transition to turbulence was subcritical
\cite{daviaud1992_prl,dauchot1995_pf} and identified the dependence on the Reynolds number (based on the velocity of the moving wall) of the minimal energy for a perturbation to trigger transition
(Table \ref{tab:couette} provides a summary of past experiments).
All of these were confined to the non-MHD case so it
remains unknown how these mechanisms may be altered in an external magnetic field.\\ 
MHD experiments were instead dedicated to the Taylor-Couette (TC) flow,
where the sheared layer of fluid is confined between concentric rotating cylinders
 rather than between infinite planes \cite{rudiger2018_pr}. 
This simplification however
comes at the cost of new families of centrifugal instabilities associated to the rotation \cite{drazin95}, and replaces one of the infinitely extended direction by a periodic one.
Aside of the fundamental interest, one of the great motivations for studying the Couette and TC flows
arises from accretion disks galaxies and the interior of stars, where magnetic fields may exist.
The phenomenologies of the MHD Couette and TC flows are also
 relevant to engineering problems, in particular when handling liquid metals \cite{patouillet2018_iopcs}.
However, while the MHD TC
problem has been extensively studied because of its direct astrophysical relevance \citep{balbus1998_rmp, seilmayer2012_prl, stefani2006_prl},
the plane MHD Couette flow, has been mostly left aside. Despite the laminar MHD Couette flow being one
of the textbooks classics \cite{moreau1990}, two papers have investigated its linear and nonlinear
stability to find it linearly stable to arbitrarily high Reynolds number \cite{kakutani1964_jps,nagata1996_jfm}, and no
experiment on the plane MHD Couette flow has ever been conducted to our knowledge.
Some of the reasons behind this gap are perhaps found in the additional challenges associated to
MHD experiments. First, the whole flow must be pervaded by a magnetic field as spatially
homogeneous as possible. Secondly, MHD experiments require
electrically conducting fluids such as liquid metals, whose opacity make direct flow visualisation
impossible. While indirect methods based on electric potential velocimetry provide partial visualisations, their application to the MHD Couette flow would be extremely difficult \cite{bpddk2017_ef}.\\
Despite these challenges, the plane configuration makes it possible to align an external magnetic field
 with velocity gradients rather than with the flow (as in most MHD TC experiments). The
shear may then oppose the tendency to two-dimensionality that results from the Lorentz force
\cite{sm82,pk2014_jfm} to produce non-trivial effects: On the one hand, an external magnetic field may
suppress instabilities in shear flows \cite{zikanov2014_amr}, and Joule dissipation provides and
extra mechanism to dissipate energy at all scales efficiently \cite{davidson2001}. On the other hand,
the change in flow structure, turbulent spectrum \cite{eckert2001_ijhff} and energy transfer mechanisms
\cite{bpdd2018_prl} incurred by the Lorentz force may suppress the dissipation mechanisms themselves \cite{pk2017_prf}. Hence, the mechanisms governing the transition to turbulence and fully developed
turbulence in the Couette flow may turn out to be very different in the MHD framework, but
remain unexplored to date.\\
We propose a solution to break the experimental deadlock based on the observation that strong MHD effects can still be
obtained with fluids of much lower conductivity, provided the loss of conductivity is compensated by
high magnetic fields. To maintain the same level of electromagnetic force,
 the Joule time $rho/\sigma B^2$, built with magnetic field $B$, the fluid's
density $\rho$ and electric conductivity $\sigma$ needs do be conserved. For example, producing
the same level of electromagnetic force in transparent electrolytes as in liquid metals that are
typically $10^4$ times more conductive but 5 times heavier, requires fields about 50 times
larger. The low conductivity however implies that the magnetic field induced by the flow is orders of
magnitude smaller than the externally applied one (unlike in astrophysical problems) so the field
is imposed and does not vary under the action of the flow \cite{roberts1967}.
Only a few experiments have taken advantage of this idea to date. The earliest might be
by \cite{andreev2003_pf} who visualised magnetoconvective patterns applying the shadowgraph method to
sulphuric acid, followed by LASER-based visualisations of the wake of an
obstacle in a magnetic field by \cite{andreev2013_jfm}. However, the first fully quantitative mapping of flow velocities was
recently achieved in a configuration mimicking the rotating magnetoconvection in the liquid core
of the Earth \cite{apbds2016_rsi}.
Being quantitative, the visualisation technique pioneered in electrolyte experiments offers an
opportunity to transpose recently developed optical methods
of pressure-PIV \cite{kat2012_ef, vanoudheusden2013_mst}. These methods take advantage of the precise mappings of the velocity fields to
numerically reconstruct the pressure field. Implementing this technique on MHD flows for the first
time involves mostly dealing with the term representing the Lorentz force in the governing equations.
\\
In this context, we set out to build an experiment providing a good approximation to the
plane MHD Couette flow  and to obtain precise mapping for the velocity and pressure fields
in both the laminar and  turbulent states for the first time, based on the PIV and Pressure PIV techniques.
The large field needed to encompass the entire flow shall be obtained by means of the stray field of a
large superconducting magnet.
In the longer run, the facility
and set of techniques developed are expected to open the way to full quantitative studies of the
transition to turbulence and the fully developed turbulence in the plane MHD Couette flows.
We shall first recall the equations governing the MHD Couette flow in the Low-Rm approximation (section \ref{sec:couette}), before
providing the technical details of the rig (section \ref{sec:experiment}). We then show how the classical PIV and pressure
PIV techniques are adapted to the MHD problem and the specific constraints of the rig (section \ref{sec:measurement}). Finally we provide
early measurements of pressure and velocity fields obtained with the setup, as a premise to a full-blown study of the transitional regime between laminar and turbulent flows (section \ref{sec:results}).
\begin{table*}
\centering
	\caption{Characteristics of existing and present Couette Flow experiments. $L_x-$ streamwise length of the domain, $L_y-$ width, $h-$ height, $U_0$ wall velocity. 
}
	\label{tab: biblio couette exp}       
	\begin{tabular}{ccccccccc}
		\hline\noalign{\smallskip}
 & Fluid & $L_x \times L_y$ /m$^2$ & $h$/mm & $L_x / h$ & $L_y / h$ & $Re$ Range & Lower $Re_c$ & $U_0$ /m.s$^{-1}$ \\
 		\noalign{\smallskip}\hline\noalign{\smallskip}
\cite{dauchot1995_pf} & water & 1.3 $\times$ 0.254 & 7 & 186 & 36 & $300-500$ & $325\pm5$ & $0.04-0.08$ \\
\cite{daviaud1992_prl} & water & 1.3 $\times$ 0.254 & 7 & 186 & 36 & $300-500$ & $370 \pm10$ & $0.04-0.06$ \\
\cite{tillmark1992_jfm} & water & 1.5 $\times$ 0.36 & 10 & 150 & 36 & $345-950$ & $360 \pm10$ & $0.03-0.1$ \\
\cite{boniface2017_epl} & water & 1.5 $\times$ 0.05  & $10-125$ & $12-150$ & $0.4-5$ &  $5-1.5\times10^3$ \\
\cite{boniface2017_epl} & water & 1.5 $\times$ 0.25 & $10-125$ & $12-150$ & $2-25$ & $5-1.5\times10^3$\\
Present			& ${\rm H}_2{\rm SO}_4$ (30 \%)& 1.2 $\times$ 0.38& $10-100$ &12 $-$ 120& 3.8 $-$ 38 &	250 $-$ $1.2\times 10^4$ && \\
		\noalign{\smallskip}\hline
	\end{tabular}
\label{tab:couette}
\end{table*}
\section{The plane MHD Couette flow}
\label{sec:couette}
\begin{figure*}
\centering
\begin{tabular}{cc}

\parbox{0.5\textwidth}{
\includegraphics[width=0.5\textwidth]{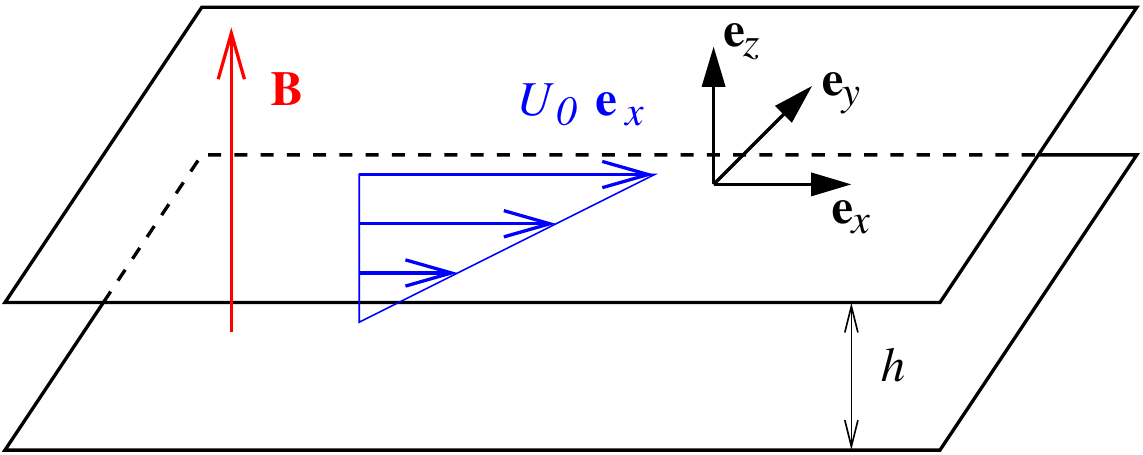}} &

\parbox{0.5\textwidth}{
\includegraphics[width=0.45\textwidth]{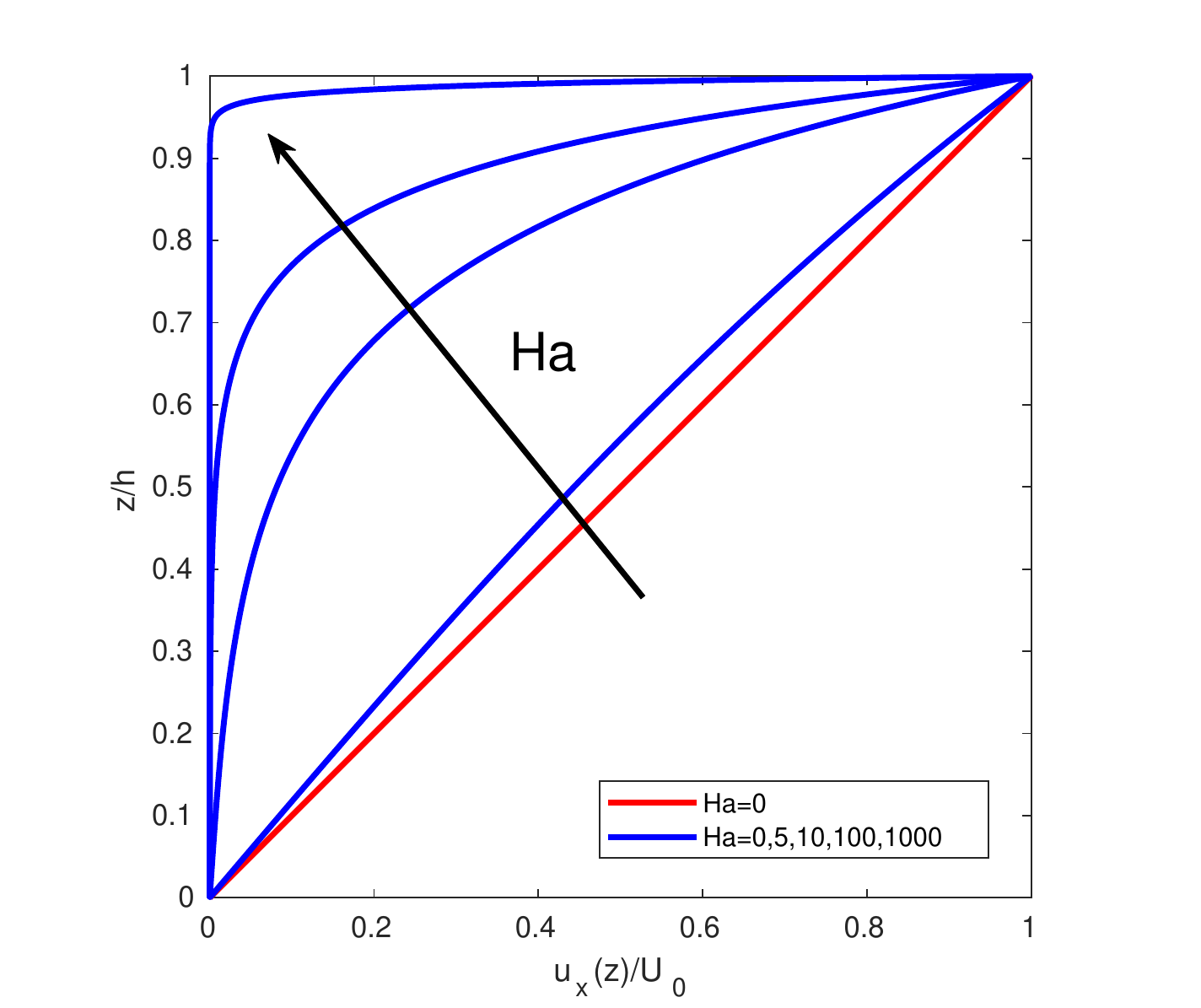}}\\
\end{tabular}
\caption{\label{fig:couette}Left: Generic configuration of the Couette flow, with top wall moving at velocity $U_0$, {fixed bottom wall, and both walls spaced by $h$}. For the MHD Couette flow a {magnetic field, of flux density} $\mathbf B$ is applied.
Right: streamwise velocity profiles for the non MHD ($Ha=0$) and MHD Couette problem. As the electromagnetic effects increase, momentum becomes concentrated in an ever thinner layer near the moving wall.}
\end{figure*}

The ideal configuration of the magnetohydrodynamic Couette flow with a transverse magnetic field is represented in figure \ref{fig:couette}: a Newtonian, incompressible fluid of density $\rho$, electric conductivity $\sigma$, viscosity $\nu$, is confined between two parallel infinite planes at $z=0$ and $z=h$, 
and pervaded by a homogeneous {magnetic flux density} $B\mathbf e_z$. The bottom wall at $z=0$ is static while the 
top wall moves at constant velocity $U_0\mathbf e_x$. For flows of moderate intensity at the scale of the laboratory, the magnetic field induced by fluid motion is negligible compared to the externally 
imposed field, and the system can be described within the 
low-$Rm$, quasi-static approximation \cite{roberts1967}. Physically,  
this approximation can be understood as taking into account the
current induced by the motion of the conducting fluid in the magnetic field, but neglecting the magnetic
field induced by this current.  The low-$Rm$ quasi-static equations are most conveniently expressed in 
terms of the velocity $\mathbf u$, pressure $p$, current density $\mathbf J$ and electric potential $\phi$. These 
include the momentum equation:
\begin{equation}
(\partial_t+\mathbf u\cdot\nabla) \mathbf u+ \nabla p=\frac1{Re}\left( \Delta \mathbf u + Ha^2\mathbf J\times \mathbf B\right), \label{eq:nsqs}
\end{equation}
Ohm's law
\begin{equation}
\mathbf J=-\nabla \phi+\mathbf u\times \mathbf B,
\label{eq:ohm}
\end{equation}
and the conservation of mass and charge:
\begin{eqnarray}
\nabla\cdot\mathbf u&=&0,\\
\label{eq:cont}
\nabla\cdot\mathbf J&=&0.
\label{eq:charge}
\end{eqnarray}
Here the equations have been written in non-dimensional form, choosing $h/U_0$ and $\rho U_0^2$ as reference time and pressure. The problem is governed by two non-dimensional parameters, the usual Reynold number $Re=U_0h/\nu$ as well as the Hartmann number $Ha=Bh(\sigma/\rho\nu)^{1/2}$, the square of which measures the
ratio of Lorentz to viscous forces.  A third number, the interaction parameter can be formed as $N=Ha^2/Re$, to measure the ratio of Lorentz to inertial forces.
At both planes, the velocity field satisfies a no-slip, impermeable boundary condition:
\begin{equation}
\mathbf u(z=0)=0, \qquad \mathbf u(z=0)=\mathbf e_x.
\label{eq:bc}
\end{equation}
In the laminar regime, an exact analytical solution to the system (\ref{eq:nsqs}-\ref{eq:charge}) subject to boundary conditions (\ref{eq:bc}) exists:
\begin{equation}
\mathbf u(z)= \frac{\sinh(Ha z)}{\sinh Ha}\mathbf e_x, \qquad \nabla p=0, \qquad \nabla \phi=0.
\label{eq:lam_couette}
\end{equation}
The Couette solution is represented in figure \ref{fig:couette}. As electromagnetic effects 
are increased (Higher field, higher value of $Ha$), the profile presents sharper gradients in the 
vicinity of the moving wall where a Hartmann boundary layer of thickness $Ha^{-1}$ develops. 
Interestingly, just like the Hartmann velocity profile \cite{moreau1990}, the MHD-Couette profile does not 
depend on the electric boundary condition at the 
walls.%

\section{The MHD plane Couette flow experimental facility}
\label{sec:experiment}
\subsection{Principle and design constraints}
The principle of the experiment is to model the top moving wall by means of a mechanically entrained tape of width $L_y$ and length $L_x$, placed in a rectangular fluid container. 
As for the infinitely extended geometry of the theoretical MHD-Couette flow, the control parameters remain the Reynolds number $Re$ and the Hartmann number $Ha$. 
The two 
aspect ratios $L_x/h$ and $L_y/h$ account for the finite streamwise length $L_x$ and width $L_y$ of 
the experimental fluid domain. In the limit where both are very large, the rig becomes an ever better
 approximation to the infinitely extended Couette flow, for which these ratios cease 
to be relevant.\\
The challenge in our endeavour is the requirement to visualize the flow patterns: this implies that the
working fluid must be both transparent and electrically conducting, as well as Newtonian. Since
transparent fluids have very weak electrical conductivities ($10^4$ times lower than for a liquid metal 
at best), but
also densities about 5 times lower, magnetic fields about 40-50 times higher than in a liquid metal
 must be used to be able to conserve the timescale of the Lorentz force $\rho/\sigma B^2$ and attain 
interaction parameters corresponding to a significant Lorentz force. This implies using high magnetic 
fields, which are not normally available in large enough volumes to accommodate large fluid dynamics 
experiments. These
conflicting constraints only leave a very narrow margin for the design of the experiment. Considering
the prohibitive cost of a bespoke magnet providing a homogeneous magnetic field over a wide surface,
 we adopted sulphuric acid concentrated at 30 \% mass
(of viscosity {$\nu=2.06\times10^{-6}$ m$^2$/s}, density $1.250\times10^3$ kg/m$^3$, and electric 
conductivity 83 S/m at 25$^oC$ \cite{darling64}) as the working 
fluid and used a large 4 T superconducting solenoidal magnet. However, size 
constraints on the experiment imply using the magnet's stray field which is limited to about 1 T. 
%
\subsection{General description of the experimental rig}
\begin{figure*}
        \centering
        \includegraphics[width=\textwidth]{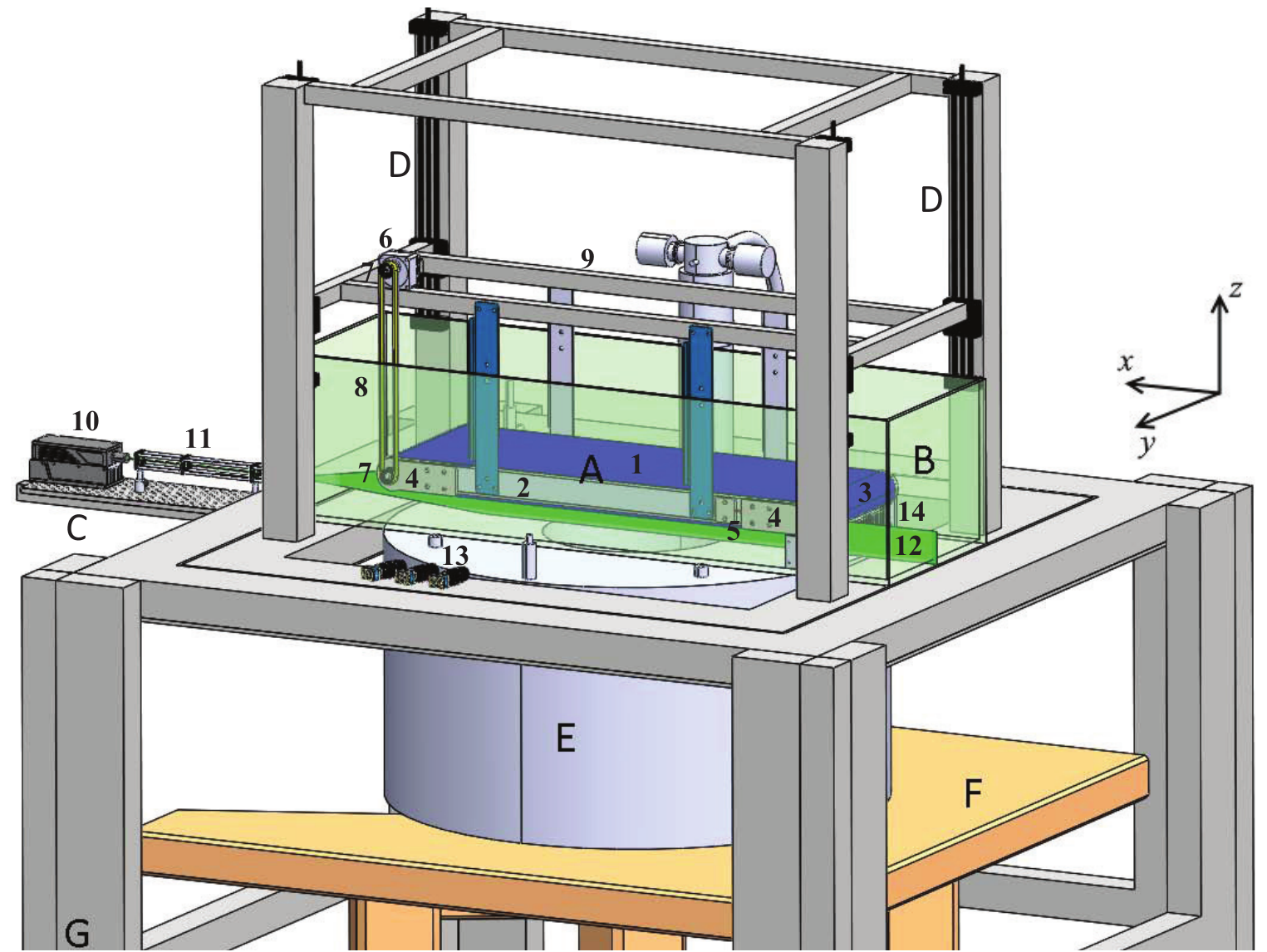}
        \caption{Sketch of the complete experimental apparatus on the left panel: A. Conveyor, B. PMMA vessel, C. PIV System (Laser, Optical system and cameras), D. Sliding mounts, E. Magnet, F. Wooden frame, G. Aluminium frame.  1. Polyurethane belt, 2. Stainless steel frame, 3. Rollers, 4. Bearing plates, 5. Pressure screw, 6. Motor, 7. Stainless steel pulleys, 8. Polyurethane timing belt, 9. Conveyor frame , 
10. Continuous Laser, 11. Optical system, 12. Laser sheet, 13. Cameras, {14. Perturbation grid (behind the LASER sheet)}. The motor pulls the polyurethane main belt so that the flow is directed from the right to the left. The external containment elements are not shown, for clarity. 
}
        \label{fig:overview_experiment}
\end{figure*}

The setup consists of a rectangular transparent vessel filled with sulphuric acid {(labelled B on figure \ref{fig:overview_experiment})}. The flow
is driven by a belt {(1)} immersed in the vessel and the Couette flow is generated between
the moving belt and the plane bottom of the vessel. The distance between them $h$ is
adjustable. The magnetic field is generated by a cylindrical solenoidal magnet {(E)} 
positioned flush a few mm under the bottom of the vessel, so the maximum stray field pervades the fluid. 
A large aluminium NORCAN\texttrademark-frame {(G)}
supports the vessel and the conveyor {(A) actioning the belt} in this position. 
The system is enclosed in 
a safety container {(not shown on figure \ref{fig:overview_experiment} for clarity)} 
whose purpose is to contain possible acid leaks and to keep the fluid
domain in the darkness that is necessary for the optical measurements. These are used to
measure velocity in vertical and horizontal planes. A general view of the setup is visible
in figure \ref{fig:overview_experiment}, and we shall now describe each of the elements in detail.
\subsection{Fluid and Light containment}
The main fluid vessel {(B)} is made of 12 mm thick PMMA plates. The total inner dimensions are 1576 mm long, 
476 mm 
wide and 388 mm high. Plates are welded together {at their edges} with PVC. This way of joining the 
plates 
makes it possible to visualise the first few millimetres above the bottom plate. 
In standard operating conditions, it is filled with about 200 l of sulphuric acid, namely about 250 mm height, so as to prevent any risk of splashing and also ensure that the Couette flow is sufficiently far from possible surface waves that could influence it.
The main vessel and the flow entrainment system (see section \ref{sec:entrainment}) are housed in a 
plastic container fitted to the magnet geometry and whose purpose is two-fold: 
 Its safety function is to contain any possible leak of sulphuric acid and to prevent any laser reflection to outside. Its light containment function is to minimise the amount of light entering the investigated fluid area, thus maximising contrast for the best possible PIV measurements.
The top container is connected through plastic pipes to a second PVC container made of 6 mm 
thick glued, grey PVC panels and placed underneath the magnet inside the wooden frame supporting the magnet.
An intermediate tube links the upper part to the lower part to protect the magnet bore. It is made with
1.5 mm thick PVC sheet and is inserted inside the bore.\\
The container where the transparent tank is introduced was made with 5 mm transparent folded and welded 
PETG panels. 
The bottom panel was covered with a very thin black sheet. 3 mm thickness black PVC plates were screwed onto the lateral panels of the PETG tank to achieve light containment.   
Two Perspex windows were fitted to the front and side PVC panels for access to optical elements and visualisations in vertical laser planes.
\subsection{Flow entrainment \label{sec:entrainment}}
The flow is driven by a conveyor {(A)} fitted with a moving belt {(1)}. As depicted in figure 
\ref{fig:overview_experiment}, the conveyor is composed of a main 
stainless steel frame “U” section {(2)} on which four bearing plates are fitted to support 80 mm diameter 
plastic rollers {(3)}, whose position can be precisely adjusted to control both the alignment and the tension of the belt. This system prevents the belt from skidding and from sagging by more than 1 mm over its 
entire surface.
The belt (made of blue polyurethane, of  length 2440 mm, width  374 mm and thickness 0.85 mm) 
is entrained by a precisely controlled electric motor from ORIENTAL MOTORS\texttrademark, regulated to achieve 
constant speed within a precision better than $0.2\%$ {(determined by the motor's 
electronic regulation)}.
The whole system slides up and down through sliding mounts {(D)} connecting the conveyor 
to the four vertical aluminium profiles located on the top of the frame and around the main vessel.\\
To control the level of fluctuations in the flow, a perturbation system {(14)} is placed 
at the inlet of the Couette flow region, just below the belt (see fig. \ref{fig:overview_experiment}). It consists of rectangular grids of various steps and height, which cover the entire section of the inlet. They are made of acid-resistant and non-magnetic stainless steel cylindrical rods of thickness 1.5 mm and separated by either 5 mm or 10 mm. They are slotted in holders fitted on the side panels of the transparent vessel, so as to be easily removed or swapped.
\subsection{Magnetic field}
\begin{figure}
	\centering
	\includegraphics[width=0.5\textwidth]{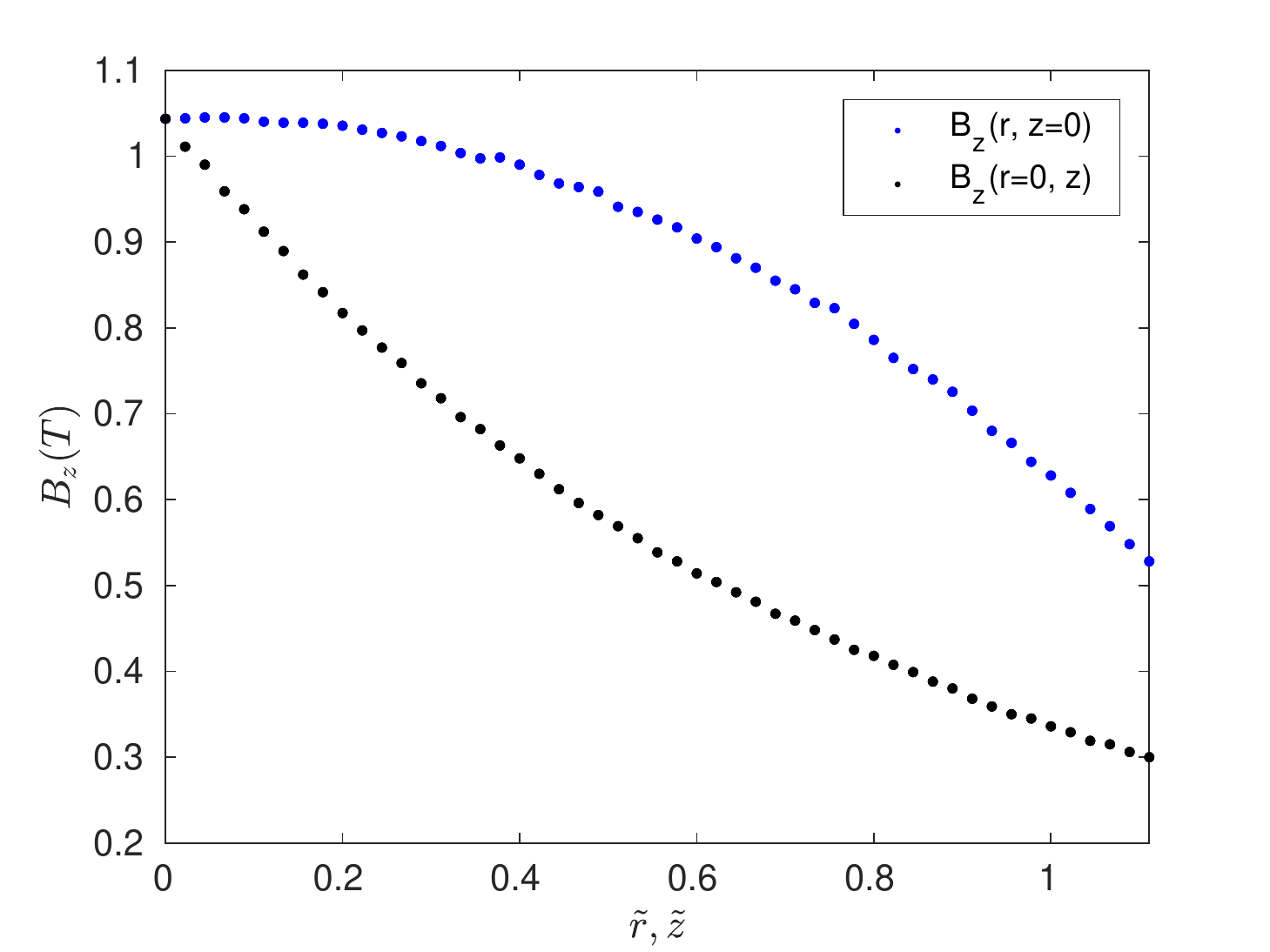}
	\caption{Profiles of axial magnetic {flux density} $B_z$ measured in the stray field at the top of the magnet along the vertical axis $\tilde{z}$  $(\tilde{z}=z/R_{\rm magnet})$ and radius $\tilde{r}$ $(\tilde{r}=r/R_{\rm magnet})$, for a field of 4 T at the centre for the magnet bore.}
	\label{fig:magnet}      
\end{figure}

The magnetic field is generated by a superconducting magnet available at the High Magnetic Field Laboratory in Grenoble (LNCMI/CNRS). It is 635 mm length, has a bore of diameter $2R_{\rm magnet}= 450$ mm and is positioned 
at 1710 mm high. It can be operated at field strengths of 0 to 4 T in its centre. The {magnetic flux density} was experimentally measured from the bottom of the transparent tank
along the centred vertical axis $z$ and along the radius $r$ at $z=0$. Both profiles are plotted in Figure \ref{fig:magnet}. 
When the field is set to its maximum value of 4 T at the centre of the bore, the {magnetic field density} 
decreases from 1.04 T at the bottom of the transparent tank ($z/h=0$) to 0.034 T at $z/h=1$ m along the vertical 
axis and from 1.04 T on the solenoid axis at $r/R_{\rm magnet}=0$ to 0.63 T at {$r/R_{\rm magnet}=1$ ($R_{\rm magnet}=225$ mm)}. Our investigated area extends 
typically from 20 mm to 60 mm above the upper surface of the magnet. The total inhomogeneity across 
the channel height varies from 15.3\% at $h=30$ mm  to 27.3\% at $h=60$ mm for $B_z$. While lateral 
inhomogeneity exceeds 30\% across the whole belt width, our area of interest 
only extends approximately 100 mm away from the centre of the magnet, with an homogeneity of around 
7\%. The time variations of the magnetic field are not measurable over timescales 
relevant to the experiment. {For comparison, in previous experiments such as FLOWCUBE \cite{bpddk2017_ef,bpdd2018_prl}, precise theoretical scaling laws on turbulence intensity and 2D/3D cutoff scales were recovered with field inhomogeneity of up to 10\%. As such the inhomogeneity in the stray field probably lies at the limit of what is acceptable for quantitative experiments in homogeneous fields.}
\subsection{Physical range of parameters}
The setup is fairly flexible and its regimes of operation are controlled by 3 physical parameters:
\begin{itemize}
\item the intensity of the magnetic field, 
\item the distance $h$ between the bottom vessel of the tank and the conveyor belt. Even though this distance could be taken as high as the vessel height, it is kept to a maximum of 60 mm in order to
keep a good vertical homogeneity of the magnetic field. Distances lower than $h=20$ mm, make measurements difficult and increase the relative error due to slight vertical movement of the belt.
\item the velocity of the belt is limited by the torque of the electric motor to 1.5 m/s and cannot be
regulated below 0.01 m/s.
\end{itemize}

These constraints imply that in theory, the setup can reach $Re$ up to approximately $7.3\times 10^4$ (for $h=0.1$) m and 
$N$ up to 0.1 at $Re=1000$ (though higher values are possible in laminar regime).
Typical operational values of these parameters are summarized in table \ref{tab:param}.
It should be noted that the belt height $h$ is not a relevant parameter of the ideal MHD-Couette problem once Ha and N are known.  
However, since the magnetic field is not perfectly homogeneous, measurements taken at the same Reynolds and Hartmann 
numbers but for different heights (and therefore different velocities) could potentially differ. This 
effect is in part mitigated by choosing the field at the moving wall as a reference for $Ha$ (and $N$).  
In the reminder of the paper, the values of the dimensionless parameters are corrected after a precise measure of $h$ on the PIV images.
\begin{table}
	\centering
	\caption{Typical dimensional parameters (height $h$, moving wall velocity $U_0$ and axial {magnetic flux density} $B_z$) and dimensionless parameters 
(Reynolds number $Re$ and Hartmann number $Ha$). The magnetic field is measured on the axis of the magnet at the location of the tape and 
therefore depends on the tape position. The interaction parameter is $N=Ha^2/Re$ {and 
the maximum values correspond to a Reynolds number of $500$ for which the flow is laminar, and below 
which the motor does not operate smoothly at $h=59.6$ mm.}}
	\label{tab:Re and N}       
	\begin{tabular}{cccccc}
		\hline\noalign{\smallskip}
		$h$ & $U_0$  &  $Re$  & $B_z$ & $Ha$ & \color{blue} $N$\\
		(mm) & (m.s$^{-1}$) &    &  (T) & & \\
		\noalign{\smallskip}\hline\noalign{\smallskip}
    30    &   0 - 1  &  0 - $1.5\times10^4$  &    0--1.05  & 0--5 & \color{blue} 0--0.05\\
	60	 &   0 - 1  &  0 -  $3\times10^4$   &    0--0.79  &  0--8.05 & \color{blue} 0--0.13\\

		\noalign{\smallskip}\hline
	\end{tabular}
\label{tab:param}
\end{table}
%
%
\section{Measurement techniques and experimental procedure }\label{sec:measurement}
\subsection{Particle Image Velocimetry (PIV) }\label{sec:piv}
\begin{figure}
	\centering
	\includegraphics[width=0.49\textwidth]{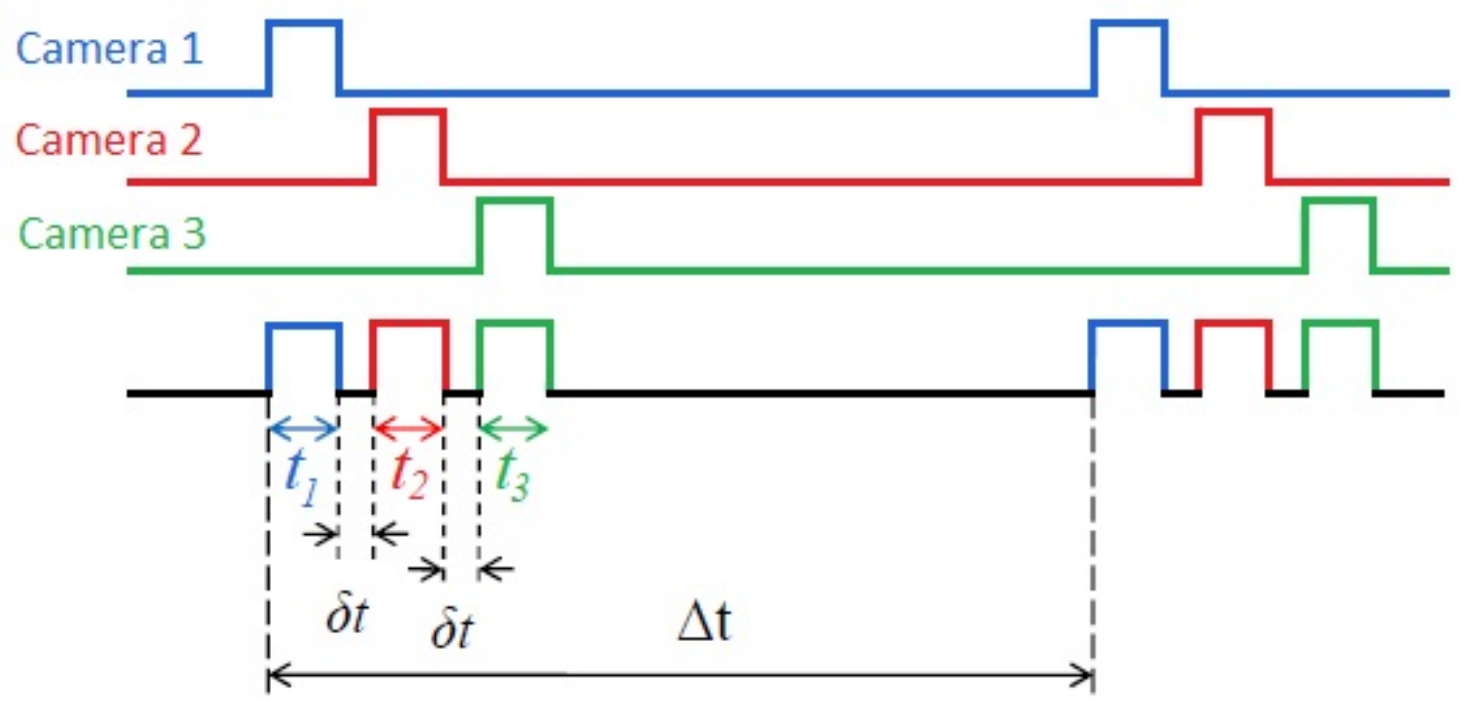}
	\caption{Measurement method with three cameras. A burst of three frames of time exposure $t_1$, $t_2$ and $t_3$ is recorded with a time between frames $\delta t$ smaller than the repetition time $\Delta t$.
	\label{fig:PIV technique}} 

\end{figure}

The principle of PIV is to seed the fluid with very small, neutrally buoyant non-inertial, highly 
reflective particles. These are then illuminated with  a  laser  sheet. Cameras record successive 
snapshots of the laser sheets and the brightness distributions of successive snapshots are then 
correlated to infer the displacement of the particles, which can be traced to a local velocity \cite{raffel_2018}. Here, deriving the pressure field from measurements of the velocity fields, requires the 
acceleration too.
Thus, three instead of two successive snapshots are required, to be able to access not only the
velocity associated to the displacement of tracers but also its time-derivative.\\
PIV measurements are performed either in a vertical plane in the middle of the belt or in horizontal planes
of adjustable height between 2 mm and $h-2$ mm. In both cases the LASER 
fires in the streamwise direction in the mid-plane of the belt, and the lens that converts the incoming 
linear beam into a plane LASER sheet is turned by 90$^o$ to obtain either a vertical or a horizontal 
sheet.\\
A continuous laser is used to emit light at 532 nm with a maximum power of 4 W {(labelled 10 on figure \ref{fig:overview_experiment})}. The laser sheet is generated with a home-made optical system {(11)}, including divergent, convergent and cylindrical lenses, to 
reach a thickness smaller than 1 mm \cite{moudjed_2013}. 
The fluid is seeded with silver-coated hollow glass sphere particles from DANTEC ($10$ $\mu$m diameter 
and $1.4 $ g.cm$^{-3}$ density). The effect of size and dimension of the particles can be estimated in te\emph{rms} of 
response time $\tau$ and relative velocity lag due to fluid acceleration $\left( u_p - u_f\right) / u_f$,
 where $u_p$ is the particle velocity and $u_f$ is the average fluid velocity in a square window of size $L$ around the particle 
\cite{raffel_2018}: 
\begin{eqnarray}
\tau& =& \rho_p\frac{d_p^2}{18\nu},\\
\frac{u_p - u_f}{u_f} &=& \frac{d_p^2}{18L^2}\left(\frac{\rho_p}{\rho} - 1\right)Re_f,
\end{eqnarray}
$\rho_p$ represents the density of the seeding particles, $d_p$ is to the diameter of the seeding particle and $Re_f=u_fL/\nu$. For the seeding particles used in the present study, the response time is estimated to about 3 $\mu$s and the relative velocity lag due to fluid acceleration is smaller than 1\%.\\
Flow visualisations are performed with Dalsa Genie Nano M1930 Monochrome cameras which offer a maximum 
continuous frame rate of 100 Hz. The specificity of this acquisition system is to use three cameras, {which, instead of recording continuously, are triggered with an external signal}. As depicted in Figure \ref{fig:PIV technique}, a burst of three frames is recorded and repeated at a lower frequency. 
The time between frames within one burst is significantly smaller than the time between two bursts: this 
technique provides sufficiently fast acquisition to record acceleration whilst avoiding the high cost 
and large datasets associated to high speed cameras. Furthermore, for flows with high average speed,
 obtaining noise-free correlations between images demands a framerate with much shorter 
period than the physical timescales of the flow. A high framerate camera would in 
fact capture a large amount of physically redundant images. Setting the period of the 3-image burst 
to a higher value that is commensurate with the flow time scales (typically a fraction of $h/U_0$, and 0.2 s for the cases at $Re\simeq 1000$--$2000$ presented section \ref{sec:results}) avoids this caveat.\\ 
The frame size is 1920x1200 pixels$^2$ and the spatial resolution is about 22.3 pixels for 1 mm of the 
visualised area.
The PIV computations are performed using Davis$^{\rm TM}$ software from Lavision$^{\rm TM}$.  
{Velocity fields are computed using the adaptive cross correlation method with starting and final correlation windows of respectively 128 and 48 pixels in size. 
Typical maximal displacements are about 25 to 30 pixels. The PIV grid is built with an overlap of 50\% 
between windows leading to 79x30 velocity vectors per PIV field at $h=3$ cm. This corresponds to one velocity vector every 1 mm in the $x$ and $y$ directions.}

\begin{figure}
        \centering
        \includegraphics[width=0.5\textwidth]{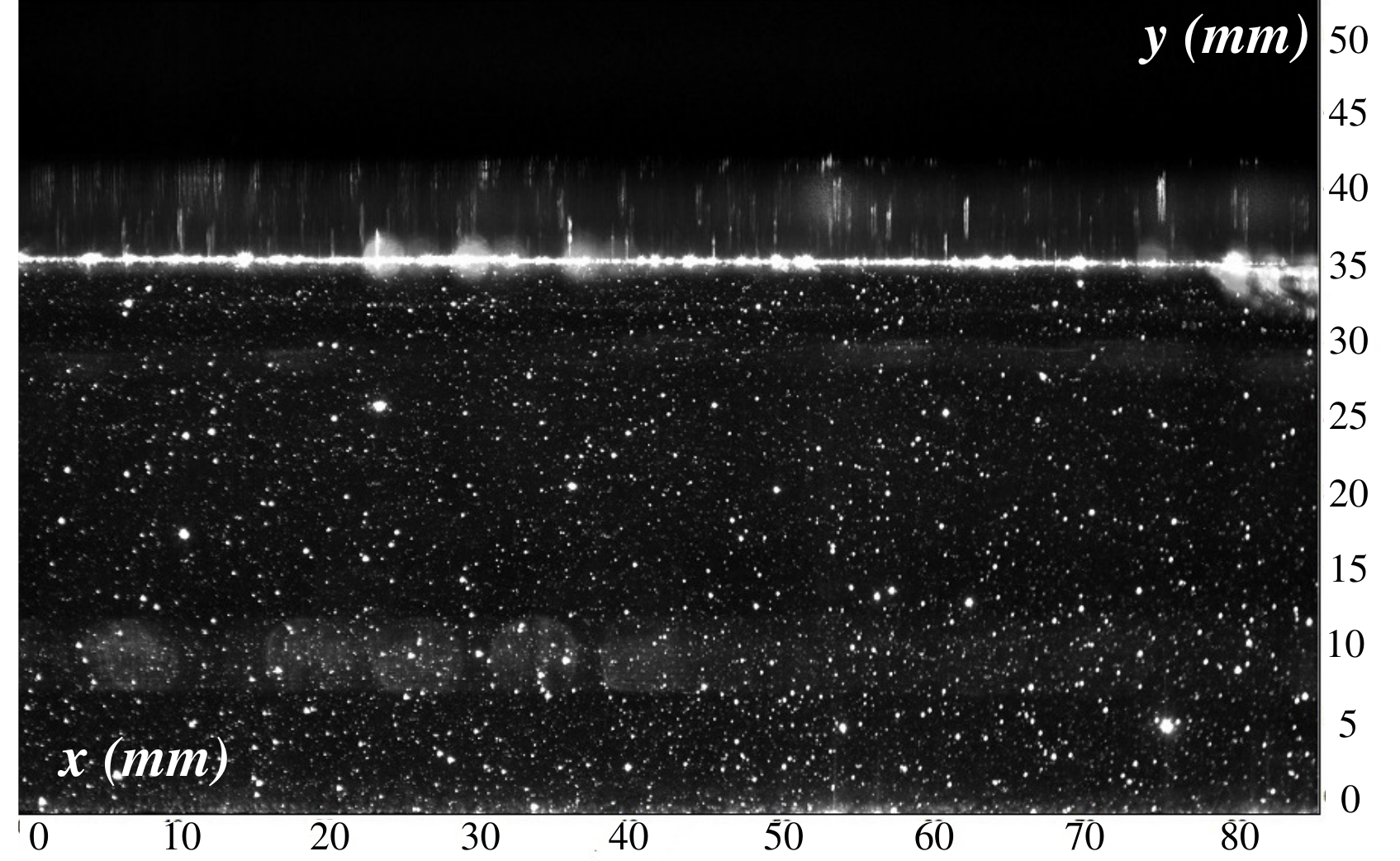}
        \caption{
Sum of corrected PIV frames acquired at the same time from all 3 cameras for measurements with the 
moving wall at $h=35$ mm. {Note that since rms of fit parameters are less than 1 pixels for all three cameras, all particles overlap perfectly on the picture.}
        \label{fig:piv_frame}}
\end{figure}

%
%
%
Measurements are calibrated by recording images of a laser printed target with
equally spaced (1 mm) dots of precisely 0.5 mm  diameter, placed
at the location of the LASER sheet (and removed during the measurements). We made sure that 
 using 3 cameras in slightly different positions did not incur any error by verifying that 
frames calibrated on each of the cameras matched each other (see Fig. \ref{fig:piv_frame}). {In the calibration procedure, the geometric parameters of a reference calibration grid (length, width and number, size and spacing between dots) are precisely known. The \emph{rms} values of the discrepancy between these reference values and the corrected calibration grid allow us to quantify the quality of image correction. Here, the \emph{rms} values are less than 1 pixel for each camera: 0.66 pixels for camera 1, 0.65 pixels for camera 2 and 0.58 pixels for camera 3. As such, the error lies just below the limit of the camera resolution.}
%
\subsection{MHD Pressure PIV}
\subsubsection{Principle}
Pressure PIV is a relatively recent technique used to derive pressure fields from velocity fields. Here, velocity fields are measured over 2D rectangular windows with the PIV technique described in section \ref{sec:piv}. The pressure field is then reconstructed solving the equations of motions (\ref{eq:nsqs}-\ref{eq:charge}) numerically. 
There are two classical approaches. In the first approach the pressure gradient is first calculated directly, and the pressure is then obtained by integration along a path originating sufficiently far from the location of interest. Though simpler, this method tends to accumulate errors along the path and may lead to inaccurate results. The second approach consists in forming a Poisson equation, using the divergence of equation (\ref{eq:nsqs}) and using mass conservation (\ref{eq:cont}):
\begin{equation}
\Delta p= -\nabla \cdot (\mathbf u\cdot\nabla \mathbf u) +N\nabla \cdot \mathbf J\times \mathbf B.
\label{eq:poiss_p}
\end{equation}
Though more complex, this method is more accurate, but requires boundary conditions for the pressure at the edge of the domain.
The main limitation of the pressure PIV technique is that to be entirely accurate, it requires knowledge of the full 3D velocity field. This can be obtained with 3D PIV techniques.
Here, the PIV system available to us only returns 2D velocity fields, so that
 some of the source terms in (\ref{eq:poiss_p}) are missing. The pressure PIV technique will only remain
accurate insofar as the missing terms remain small. 
Finally, an additional difficulty is that since we are dealing with MHD flows, the Lorentz force 
generates an extra source term in (\ref{eq:poiss_p}). We shall however see that, this term can be 
entirely expressed in terms of the velocity field and that consequently, existing pressure-PIV techniques can be  adapted to the MHD problem we consider.
\subsubsection{Implementation}
The implementation is a purely numerical task since the pressure PIV relies on the data provided by the
standard PIV described in section \ref{sec:piv}. The geometry of interest is either of the rectangular
PIV windows. For the purpose of this study, we shall focus on the vertical window (in the $x-z$ plane), only. 
The pressure field is computed 
through Poisson's equation (\ref{eq:poiss_p}). According to \cite{vanoudheusden2013_mst,kat2012_ef}, out 
of plane motion does not affect seriously the determination of the
pressure as long as gradients in that direction are smaller than in the in-plane
direction. This condition can be expected to be reasonably satisfied, at least for the mean flow, in the 
vertical plane since the main velocity gradient results from basic MHD Couette flow profile in the $x-z$
plane. 
The novelty of implementing the pressure PIV technique for the MHD
Couette flow arises from the extra source term in the Poisson equation due to the Lorentz force.
Using (\ref{eq:nsqs}) and (\ref{eq:ohm}), this term is expressed as:
\begin{eqnarray}
N\nabla_{xz} \cdot \mathbf J\times \mathbf B = N \nabla_{xz} \cdot\left(-\nabla\phi\times\mathbf {e_z}\right) + \nonumber \\
N\nabla_{xz}\cdot\left[\left(\mathbf {u}\times\mathbf {e_z}\right)\times\mathbf {e_z}\right], 
\end{eqnarray}
with $\nabla_{xz} = \frac{\partial u_x}{\partial x} + \frac{\partial u_z}{\partial z}$. 
The first term $-N\nabla_{xz}\cdot\left(\nabla\phi\times\mathbf{e}_x\right)$ cancels out exactly,
so that the in-plane divergence of the Lorentz force can be expressed as:
\begin{equation}
\nabla_{xz} \cdot \mathbf {F}_{L} = -N\dfrac{\partial u_x}{\partial x}.
\end{equation}
%
The divergence of inertial terms, by contrast, involves out-of plane components that are neglected for
the purpose of the pressure PIV technique in the $x-z$ plane. Finally the Poisson equation for the
pressure in that plane is expressed as:
\begin{eqnarray}
\frac{\partial^2p}{\partial x^2} + \frac{\partial^2p}{\partial z^2} = \nonumber \\
 - \left[\left(\frac{\partial u_x}{\partial x} \right)^2 + 2\frac{\partial u_x}{\partial z}\frac{\partial u_z}{\partial x} + \left(\frac{\partial u_z}{\partial z} \right)^2 \right] \nonumber\\
- \left( \frac\partial{\partial t} +u_x\frac{\partial}{\partial x} + u_z\frac{\partial}{\partial z} \right)\nabla_{xz}\mathbf u - N\frac{\partial u_x}{\partial x}. 
\label{eq:poissp_xz}
\end{eqnarray}

Next, boundary
conditions need to be applied. At the lower and upper (moving) walls, the z-component of the Navier
Stokes equations (\ref{eq:nsqs}) readily implies that an inhomogeneous Neumann conditions applies:
\begin{equation}
\left(\frac{\partial p}{\partial z}\right)_{z=0,1}  = \frac1{Re}\frac{\partial^2u_z}{\partial z^2}.
\label{eq:pbc_z}
\end{equation}
The inlet and outlet boundary conditions, are, by contrast not constrained by the governing equations
and are only required because in taking the divergence of the governing equations, their order has been
increased. A usual, somewhat arbitrary choice is to apply a homogeneous Dirichlet condition;
\begin{equation}
(p)_{x=x_{\rm inlet}, x_{\rm outlet}}=0.
\label{eq:pbc_x}
\end{equation}
The Poisson equation (\ref{eq:poissp_xz}) together with boundary conditions (\ref{eq:pbc_x}) and (\ref{eq:pbc_z}) are solved numerically at every timestep (\emph{i.e.} made of a burst of 3 consecutive images
to resolve the time derivatives) by means of a centred finite difference scheme of second order in
space and time. The code was validated using the non-MHD ($N=0$) analytical solution for the velocity field and the pressure gradient of 
Stuart vortices, modified to include a streamwise translation at constant velocities ranging within $v_0=$3-10 mm/s. The numerical solution was 
calculated on a grid corresponding to our PIV setup but setting $p=0$ on all boundaries to better match the infinite domain where the 
analytical solution was obtained. For $K=1.1$ (see \cite{meiron2000_jfm, oreilly2003_jfm}), the relative \emph{rms} error based on the $\mathcal L^2$ norm over the whole domain on the pressure gradient over one period remained below 0.017\%. 
\subsection{Experimental procedure}
Once the PIV system is calibrated, the perturbation grid is inserted if required, and the conveyor is 
lowered to achieve the prescribed value of $h$, 
checking horizontality and operability of the belt in the process. Next, the magnetic field is set to 
the required value. This operation is longest as ramping up the field from 0 to 4 T takes approximatively
 3 hours. The motor driving the belt is then operated to achieve the target velocity of the belt.
The belt reaches its nominal velocity in a matter of seconds. Without magnetic field a pessimistic 
estimate for the flow establishment time can be obtained from the laminar timescale for diffusion of momentum 
across the the layer 
$h^2/\nu$, \emph{i.e.}  $\sim33$ min at $h=0.06$ m. In the presence of magnetic field, momentum diffusion 
across the layer of a perturbation of size $l_\perp$ takes places over 
{$\tau_{2D}= (h/l_\perp)^2 \rho/\sigma B^2$} \emph{i.e.} $\sim 15$ s for $l_\perp=h$ at $B=1$ T \cite{sm82}.   
PIV measurements are then collected once the flow is in a statistically steady state,  for a duration of 
typically 10 min.
\section{Experimental results}
\label{sec:results}
For the purpose of identifying the effect of the magnetic field on the flow profiles, we shall use visualisations in the vertical 
plane only, without perturbation grid and for a channel heights of 32 mm and 59.6 mm, $Ha=0$ (without magnetic field) and the highest magnetic field currently 
available to us ({flux density of} 4 T at the centre of the magnet bore), for which $Ha=5$ ($h=$32 mm) and $Ha=8.05$ ($h=$59.6 mm)). A more 
quantitative analysis of the MHD Couette flow is left for future studies.
{In all cases presented here, the flow is always directed along the motion of the conveyor belt. Short PIV measurements in a horizontal plane near the belt and measurements in the vertical plane indicate that 
the flow returns above the container, and not on the side nor between the belt and the bottom of the vessel. The free surface located approx 25 cm above the bottom of the vessel showed no sign of significant surface waves so it is safe to assume that it has no detectable influence on the flow. 
Furthermore, measurements in the horizontal plane did not reveal any noticeable bias toward central or outer regions (along $y$) so
global three-dimensional effects don’t seem to play a role, at least in the regime of parameters we explore here (bearing in mind that most of the fluctuations discussed in this section are, of course, three-dimensional).}
%
%
\subsection{Velocity measurements}
\label{sec:velocities}
%
\begin{figure*}
\centering
\begin{tabular}{cc}
\parbox{0.5\textwidth}{\includegraphics[width=0.5\textwidth]{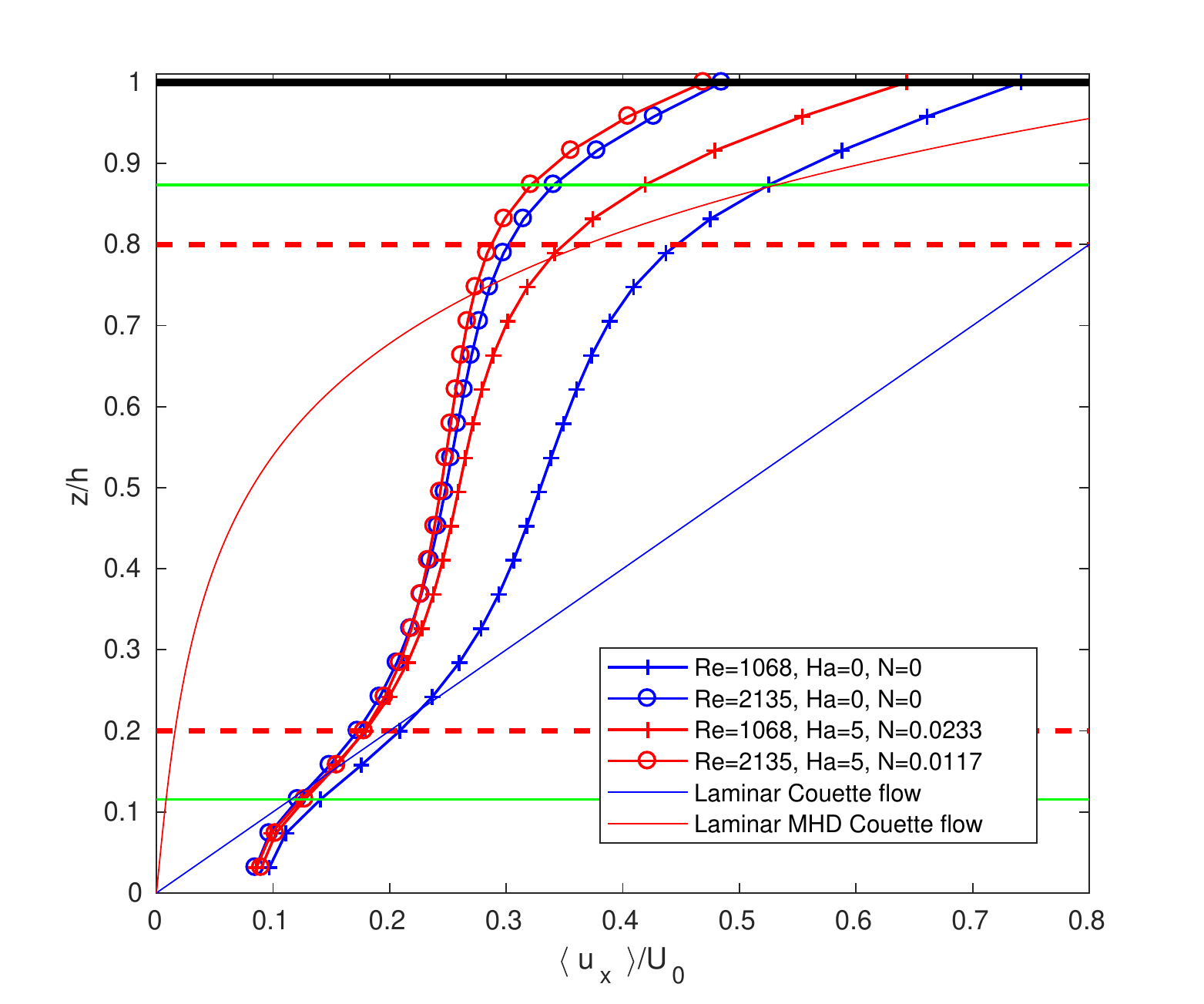}}&
\parbox{0.5\textwidth}{\includegraphics[width=0.5\textwidth]{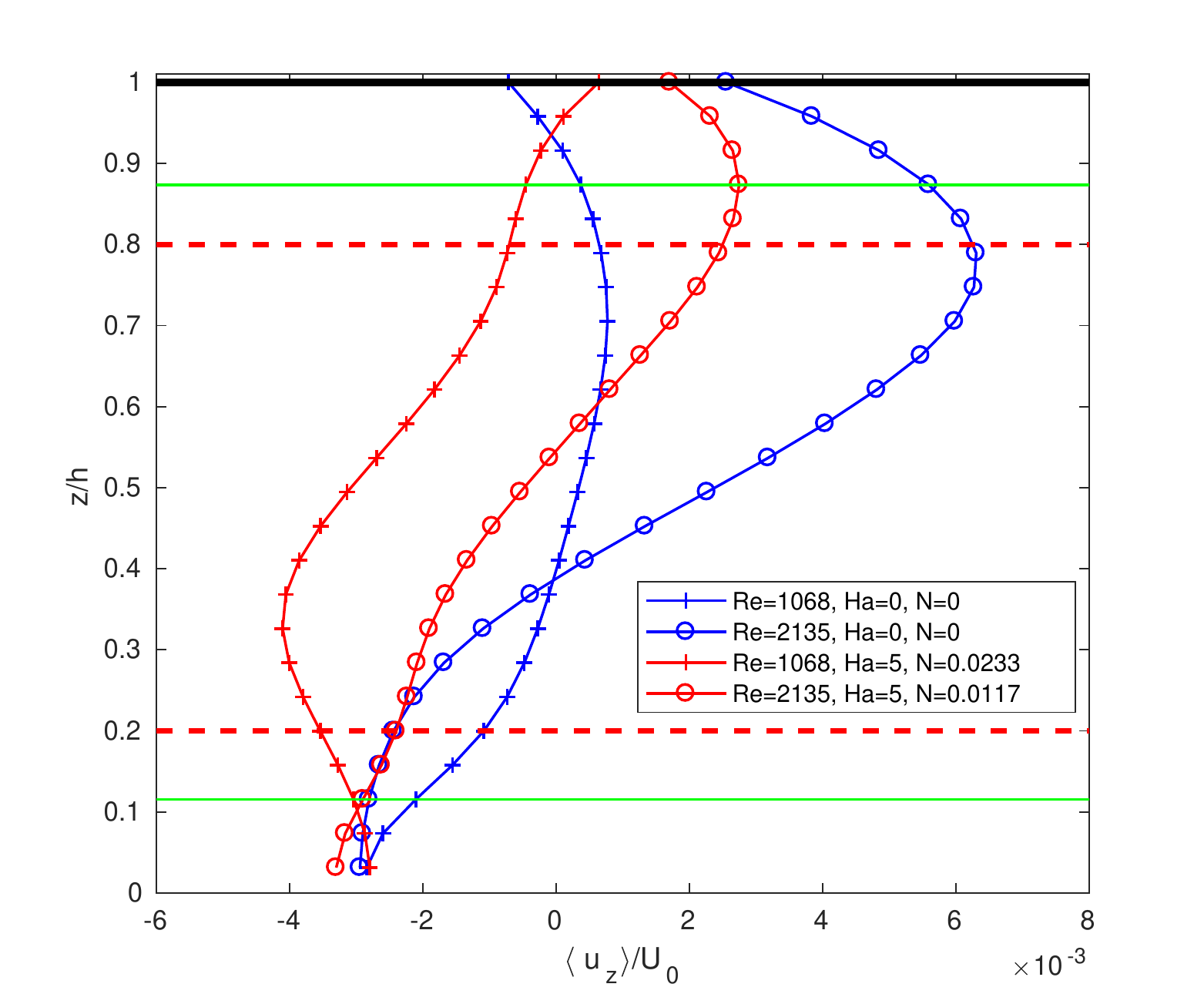}}\\
\parbox{0.5\textwidth}{\includegraphics[width=0.5\textwidth]{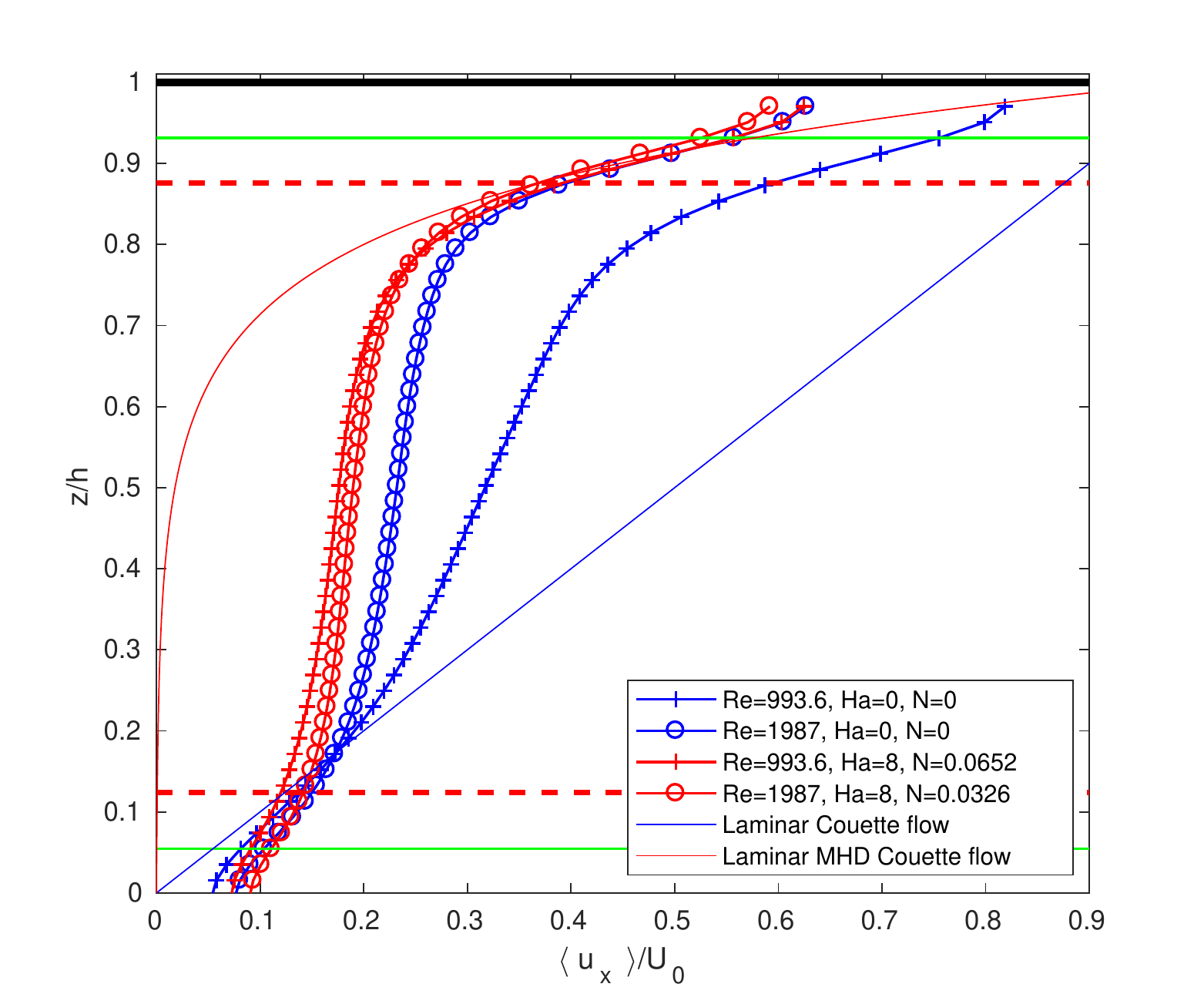}}&
\parbox{0.5\textwidth}{\includegraphics[width=0.5\textwidth]{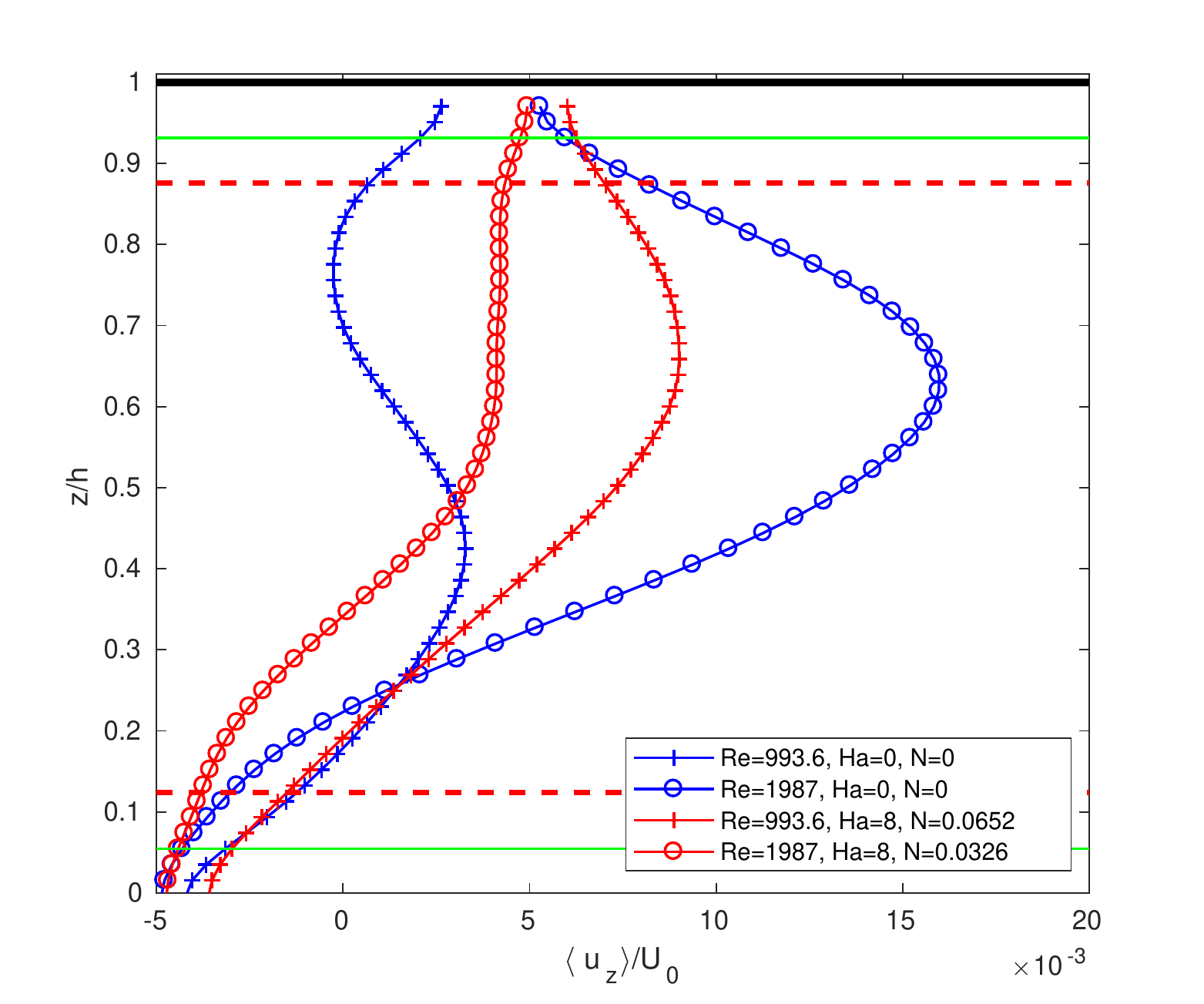}}\\
\end{tabular}
\caption{\label{fig:umean} Vertical profile of mean streamwise (left) and spanwise (right) velocities 
for several values of $Ha=0$ and $Ha=5$ (top) and $Ha=0$ and $Ha=8.05$ (bottom), with no perturbation 
grid at the inlet. The red dashed line shows the location of the Hartmann layers, while the green solid 
lines indicate the limits of reliability of the PIV data. The black line indicates the position of the 
moving wall. Dimensionally, measurements at $Ha=5$ (\emph{resp.} $Ha=8.05$) were obtained with $h=32$ mm 
(\emph{resp.} $h=59.6$ mm) and $B=4$ T at the centre of the magnet bore.}
\end{figure*}
Values of the Reynolds number between $Re=993.6$ and $Re=2135$ are considered. The profiles of mean streamwise velocity obtained with 
and without magnetic field are shown on figure \ref{fig:umean}-(left). {Dimensionally, these correspond to tape velocities of either $3.5\times10^{-2}$ m/s ($Re=993$ at $h=59.6$ mm), 
$7\times10^{-2}$ m/s ($Re=1068$ at $h=32$ mm or $Re=1987$ at $h=59.6$ mm) or $1.4\times10^{-1}$ m/s ($Re=2145$ at $h=32$ mm).} In the absence of magnetic field, the profiles show a nearly 
constant velocity in the bulk, a sharp gradient in the vicinity of the moving wall (near $z/h=1$), and a linear region near the 
bottom wall. These features are expected from turbulent Couette flows. They appear more pronounced at $Re\simeq2000$ than $Re\simeq1000$ for both channel heights, as does the reduction in thickness of both top and bottom shear layers. 
Note that missing points near $z=0$ and the absence of convergence to a value of $u_x/U_0=1$ in the vicinity of $z/h=1$ are due to limitations of the PIV setup, which cannot return valid values of the velocity in regions very close to boundaries (approx. within 3 mm). For these reasons, areas of reliable PIV data are delimited by green lines on the figures. 
Unsurprisingly, mean vertical velocities (Fig. \ref{fig:umean}-(right)) stand approximately two orders of magnitude lower
than the streamwise velocity. As the flow becomes more unstable, residual vertical
velocities due to passing perturbations develop up to a fraction 0.01-0.02  of the belt velocities. The
reason these values are not closer to zero on average will be better understood from the analysis of flow fluctuations.\\
For $Ha=5$, 8.05 and all Reynolds numbers, the mean streamwise velocity is in every point smaller than for $Ha=0$, leading to 
a significant reduction of the total flowrate. Unlike for $Ha=0$, the profiles at $Re=993.6$ and $Re=1987.6$ nearly coincide. This may be attributed to the balance between the Lorentz force and viscous friction playing a dominant role in the boundary layer near the moving wall in both cases. Indeed the thickness of the shear region is consistent with the theoretical thickness of the Hartmann boundary layer (indicated with red dashed lines on the graph). In the bulk, the velocity for $Ha=8.05$ is down to around half its value at $Ha=0$, and slightly lower at $Re=993.6$ than $Re=1987.6$, as inertial effects are weaker at lower $Re$ against the Lorentz force ($N=0.0326$ vs. $N=0.0652$). Since the effect of the field is significant, the reason both profiles differ so little for both values of Reynolds, is that with an imposed velocity at the tape, a boundary layer thickness 
imposed by the value of $Ha$, and only residual velocity in the bulk, the profile is strongly constrained. As such, it takes dominating 
inertia to break this constraint and take the flow outside the MHD regime. By contrast, for $Ha=5$ profiles at $Re=1068$ and $2135$ 
differ significantly, with barely noticeable effects of the magnetic field at $Re=2135$, for which $N=0.0117$. This suggests that the 
Lorentz significantly influences the mean velocity profiles for $N\simeq0.02$ but doesn't below this value.\\
The effect of the magnetic field on the average vertical spanwise velocity is less obvious, partly because of the
experimental difficulties in resolving their very low values, but also partly because of the inherently
erratic nature of this quantity, that is exclusively associated to large, rare perturbations. Nevertheless, the residual value being closer to 0 for $Ha>0$ than $Ha=0$, provides an early indication that the $z-$ velocity component is damped by the magnetic field.\\
%
\begin{figure*}
\begin{tabular}{cc}
\parbox{0.5\textwidth}{\includegraphics[width=0.5\textwidth]{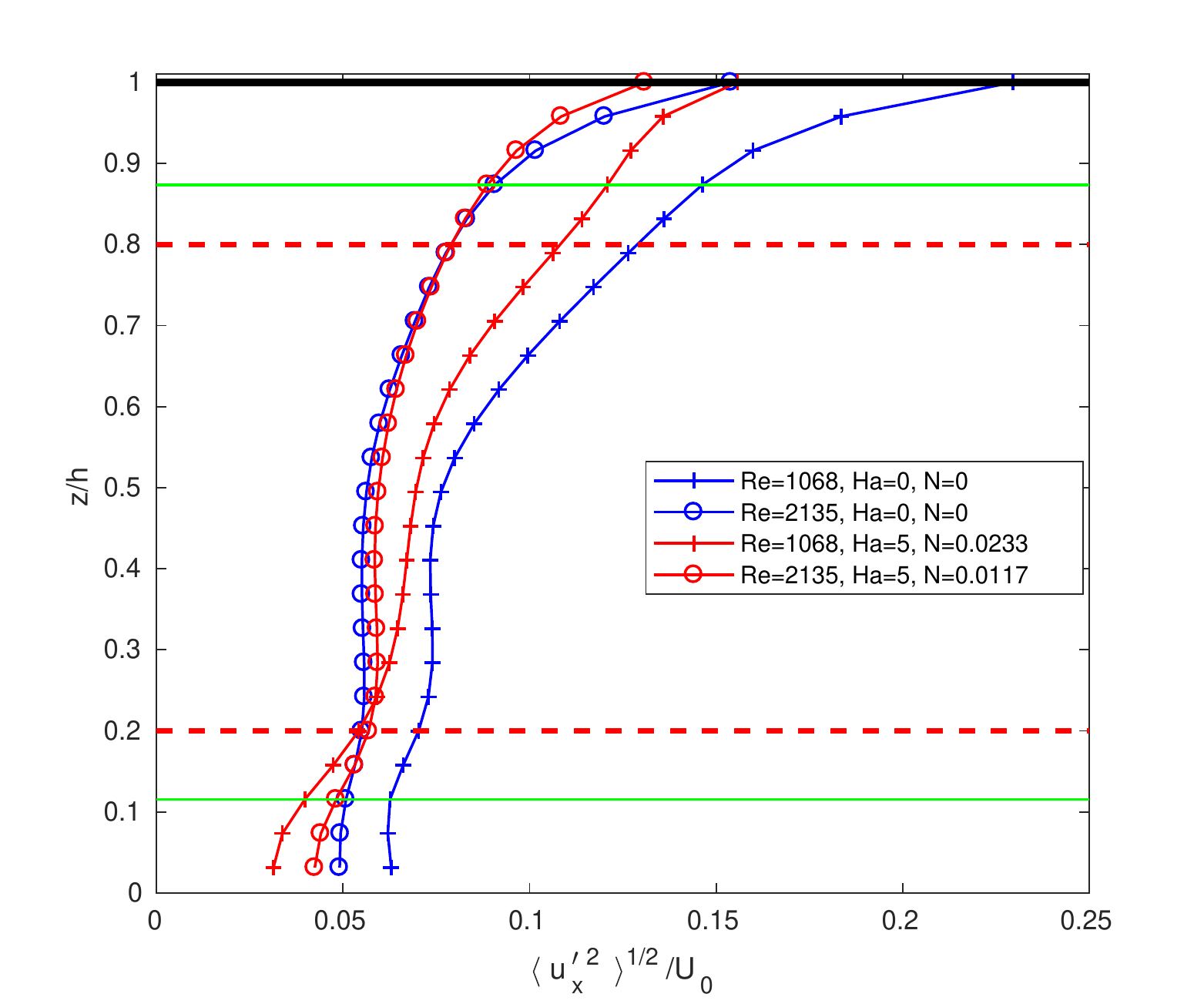}}&
\parbox{0.5\textwidth}{\includegraphics[width=0.5\textwidth]{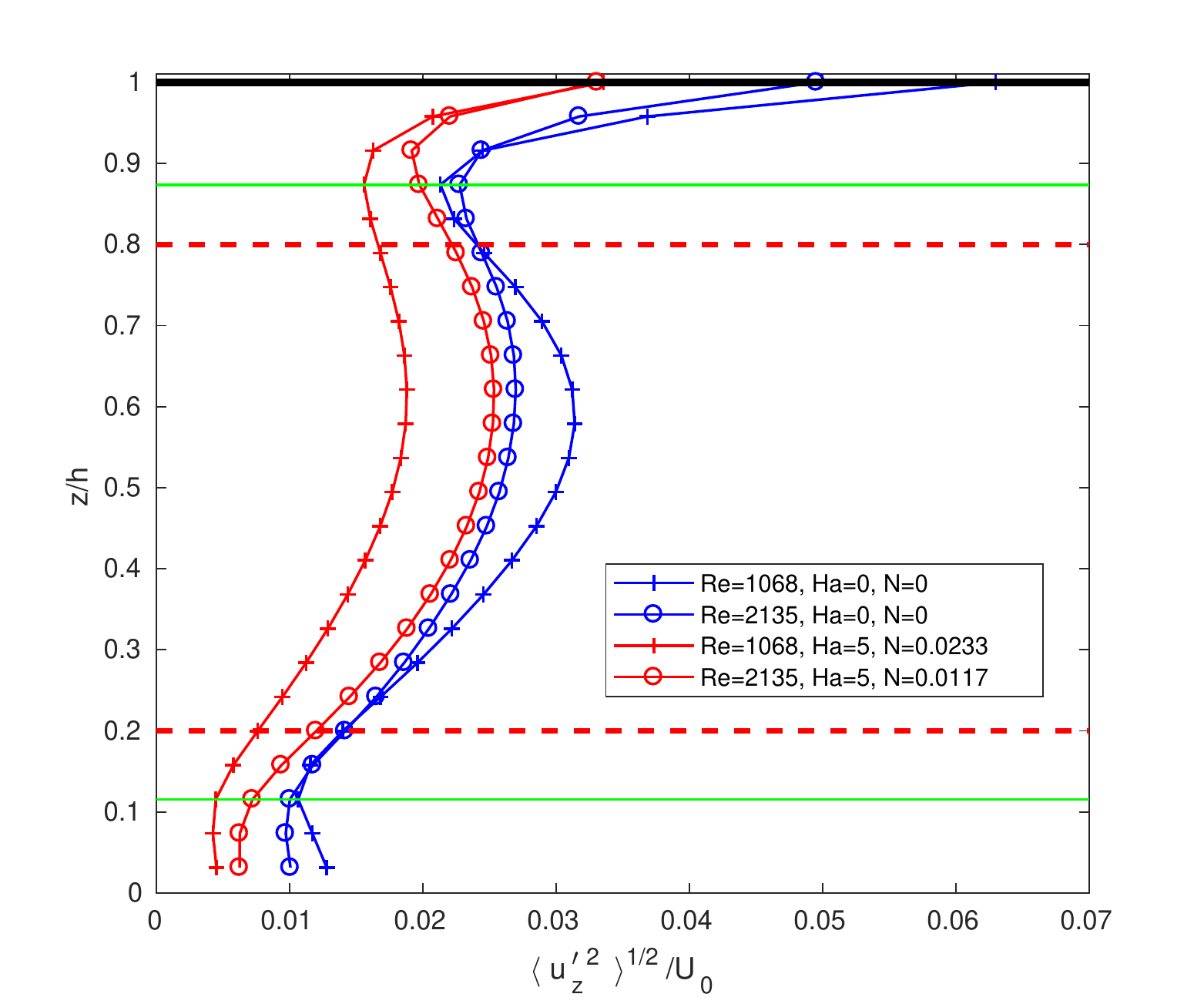}}\\
\parbox{0.5\textwidth}{\includegraphics[width=0.5\textwidth]{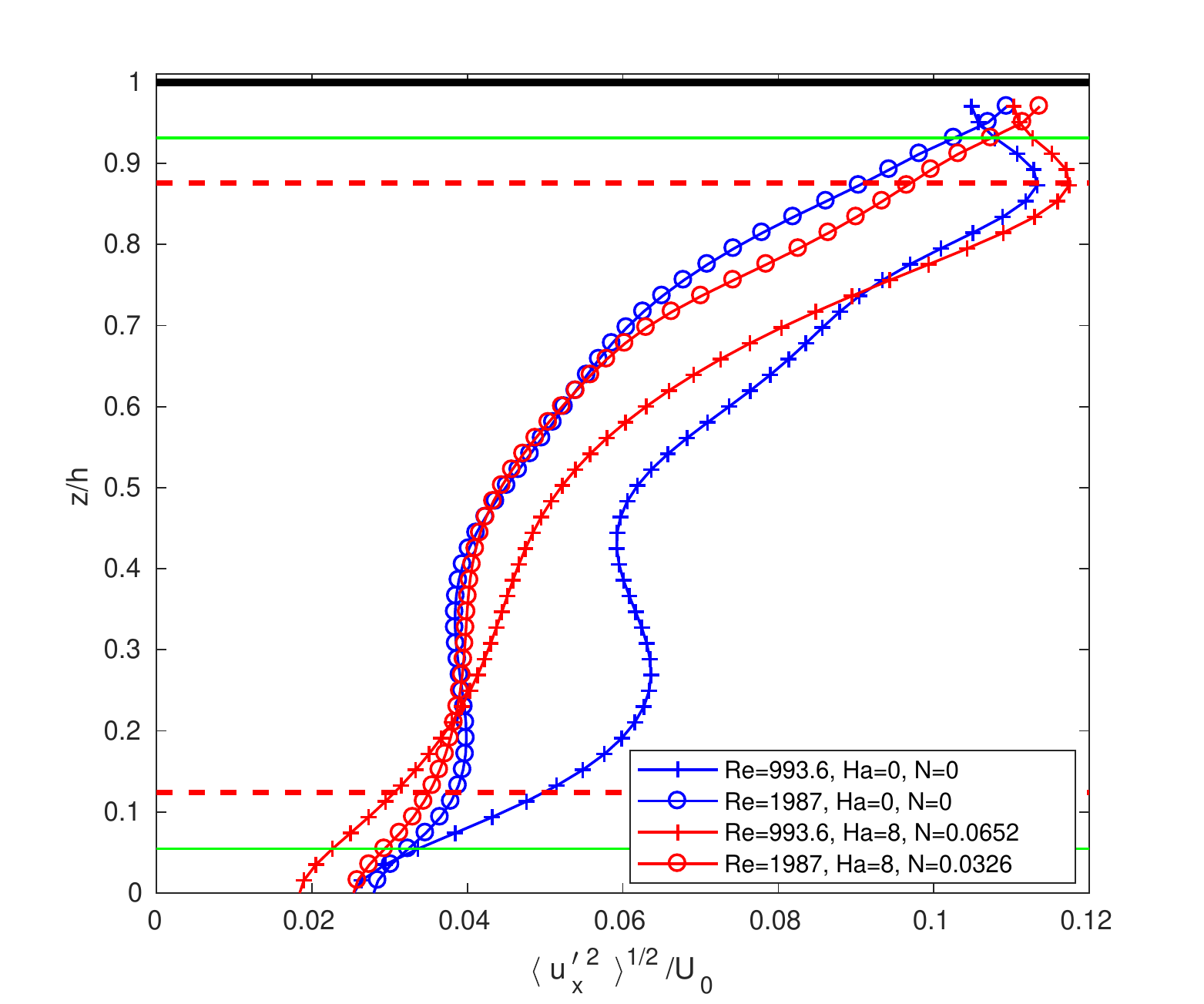}}&
\parbox{0.5\textwidth}{\includegraphics[width=0.5\textwidth]{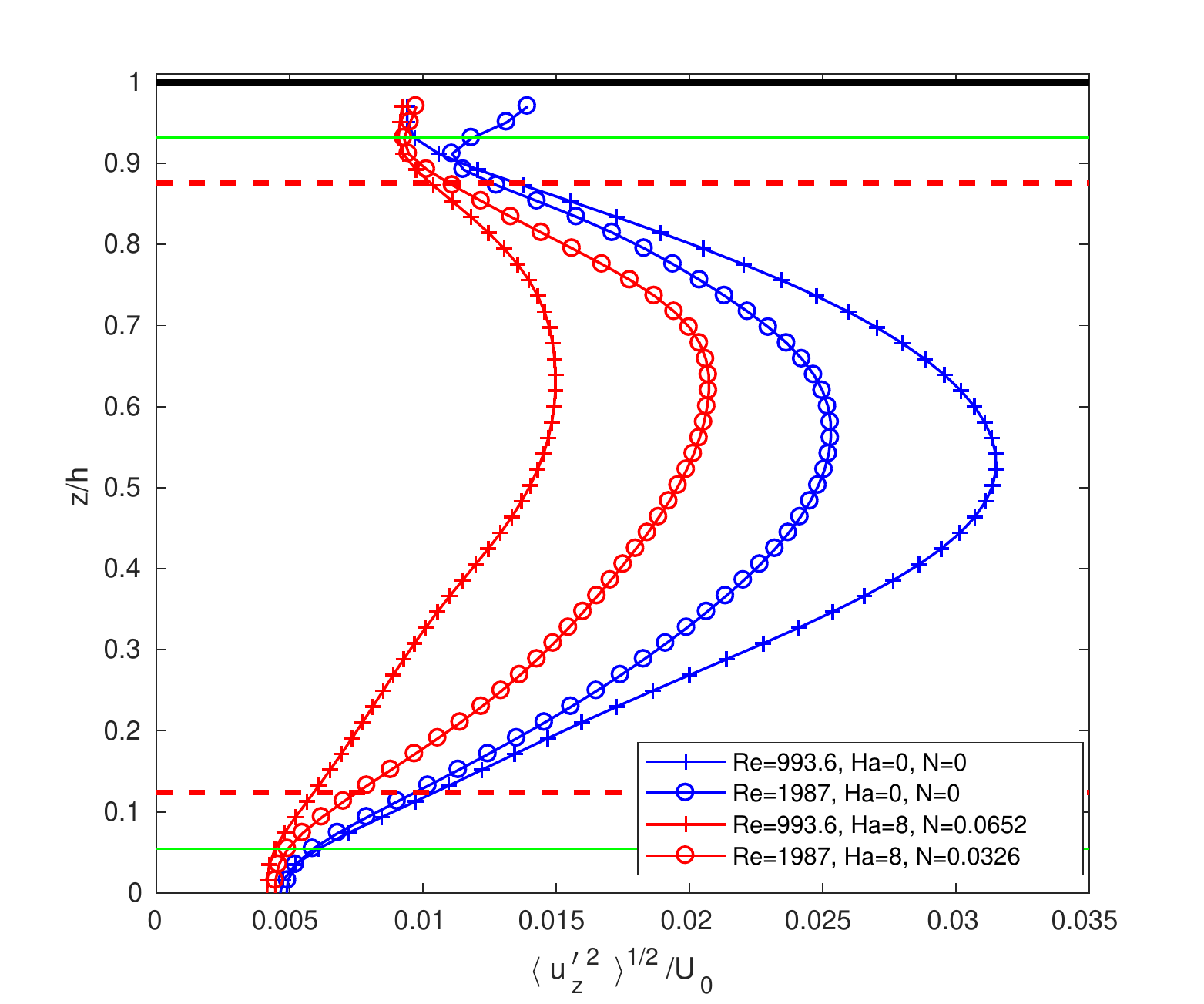}}\\
\end{tabular}
\caption{\label{fig:urms} Vertical profile of the \emph{rms} of 
streamwise (top) and spanwise (bottom) velocity fluctuations 
for $Ha=0$ and $Ha=5$ (top) and $Ha=0$ and $Ha=8.05$ (bottom), with no perturbation grid at the inlet. 
Dimensionally, measurements at $Ha=5$ (\emph{resp.} $Ha=8.05$) were obtained with $h=32$ mm  
(\emph{resp.} $h=59.6$ mm) and $B=4$ T at the centre of the magnet bore.
{The red dashed line shows the location of the Hartmann layers, while the green solid
lines indicate the limits of reliability of the PIV data.}
}
\end{figure*}
With or without magnetic field, relative streamwise velocity fluctuations (Fig.\ref{fig:urms}) 
remain mostly in a range between 0.04 and 0.12. 
The intensity of spanwise fluctuations is lower, 
around 0.01 
indicating a strong anisotropy of the underlying flow structures.
The profiles of streamwise velocity fluctuations exhibit two regions of higher  intensity, the more prominent of which is located near the top ($z/h\simeq0.8$) and the weaker one near the bottom ($z/h\simeq0.2$). These regions correspond to the passing of isolated structures. Together with the damping of vertical velocity near the walls, these explain the higher intensity of the vertical velocity fluctuations in the middle of the layer.\\
%
%
The effect of the Lorentz force is much more noticeable on the $z-$component of the velocity fluctuations: while the difference between streamwise velocity fluctuations at $Ha=5$ and $Ha=8.05$ on the one hand and their $Ha=0$ counterpart on the other is down to the level of statistical convergence of the data, 
the spanwise velocity component is clearly damped in all MHD cases, and all the more so at the interaction parameter $N$ is high. The effect is even still slightly noticeable at $N=0.0117$. Indeed, even for this value, if the ratio of the Lorentz force to inertia was evaluated using the length and velocity scales of fluctuations  (respectively the width the the region of higher transverse fluctuations $\simeq0.8h$ and the fluctuation velocity $\simeq 0.1U_0$), the corresponding parameter $N^\prime$ would lie in the region of $N^\prime\simeq0.1$, as opposed to $N\simeq0.01$. As such perturbations are significantly more sensitive to the Lorentz force than the mean flow. It is also noteworthy that the damping of the field-aligned velocity component in the presence of Hartmann walls \citep{pko2015_jfm} persists in the presence of 
strong mean shear. By contrast, this component does not vanish in periodic or unbounded geometries 
\cite{moffatt1967_jfm,schumann1976_jfm, davidson1997_jfm}. 
%
%
\subsection{Energy fluctuations}
We shall now examine more in detail the issue of statistical convergence. This issue is particularly 
apparent through the fact that in both MHD and non-MHD cases, average $z-$components of the velocity 
represent up to about 30\% of the intensity of the transversal fluctuations, when this average is 
expected to be 0. 
An element of explanation for it
is found through a closer analysis of the time-dependence of 
the total energy of the fluctuations in the measurement area, defined as
\begin{equation}
E(t)=\iint_{x,z} \mathbf u(x,z)^{\prime 2} dxdz.
\label{eq:energy_v}
\end{equation}
%
\begin{figure}
\begin{tabular}{c}
{\includegraphics[width=0.47\textwidth]{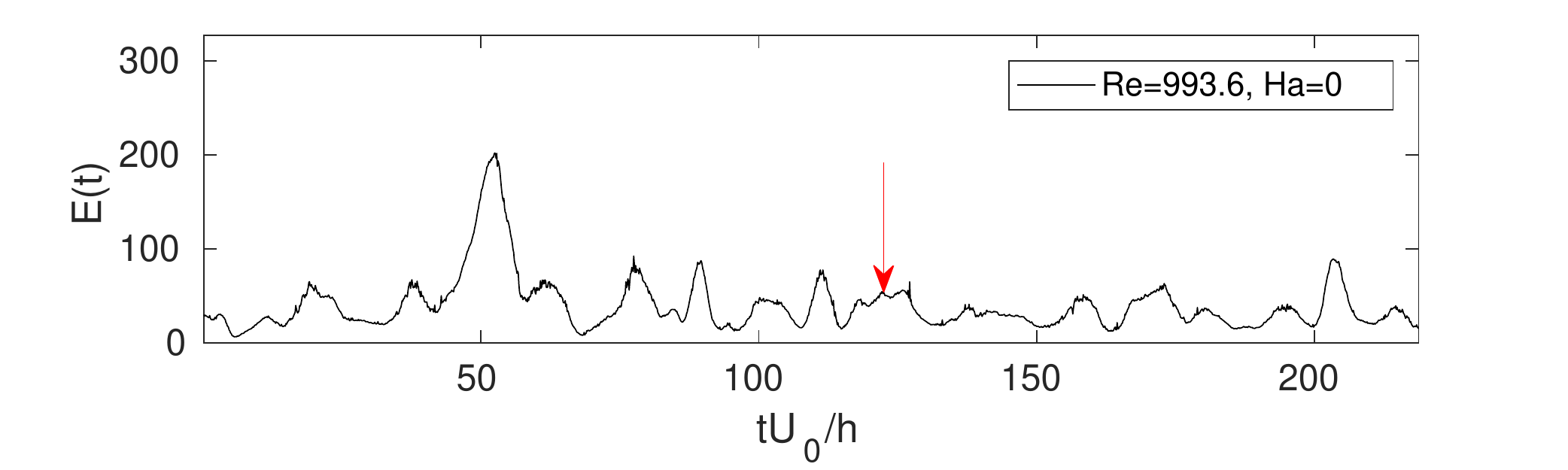}}\\
{\includegraphics[width=0.47\textwidth]{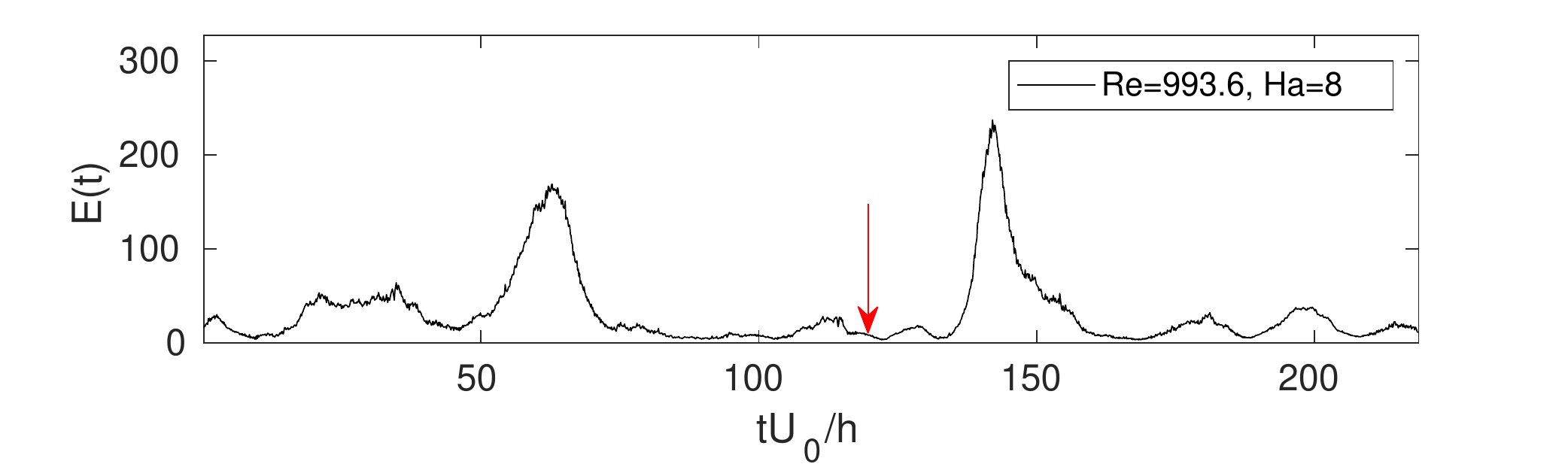}}\\
{\includegraphics[width=0.47\textwidth]{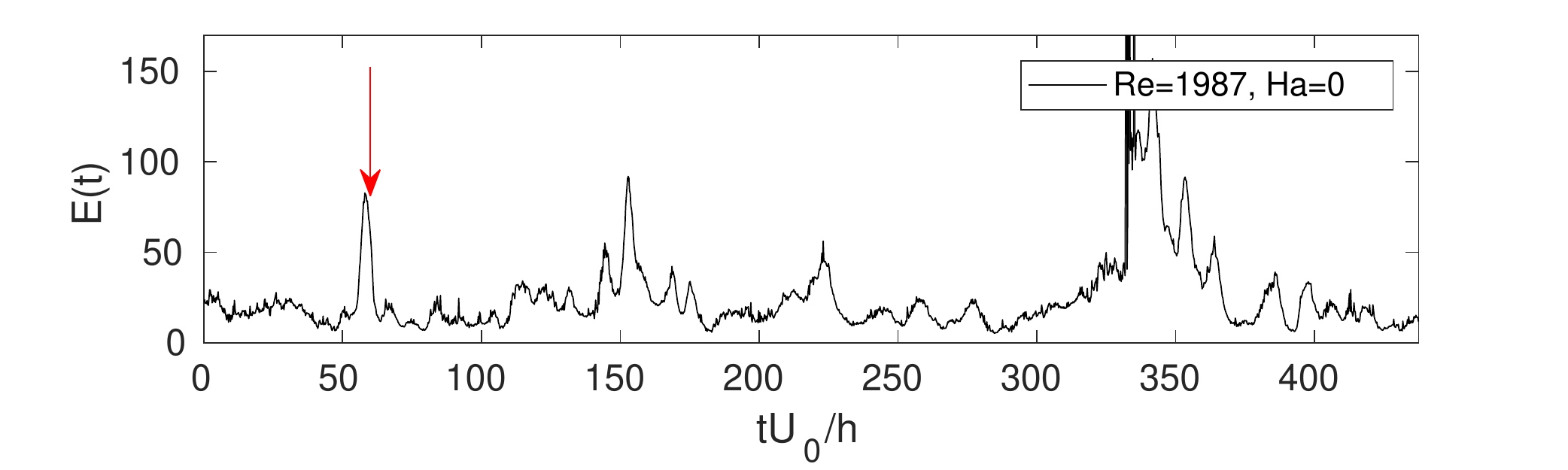}}\\
{\includegraphics[width=0.47\textwidth]{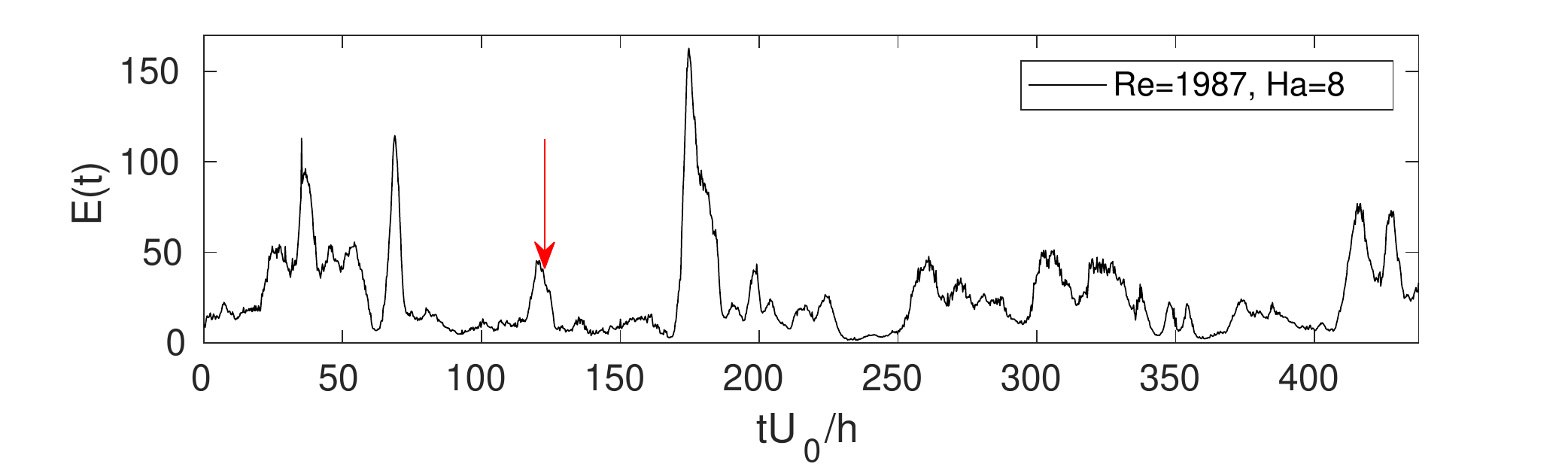}}
\end{tabular}
\caption{\label{fig:eprime_evol} Time evolution of the total energy of velocity fluctuations for $Re=1000$ and $Re=2000$ for $Ha=0$ and $Ha=8.05$ in the vertical plane. Besides the lower base level of fluctuation at $Ha=8.05$ compared to $Ha=0$, rare but intense fluctuations persist that carry most of the energy. 
Note that sharp isolated peaks are faulty images, that are filtered out for the processing of statistical quantities and spectra. Extreme events translate into peaks in the vertical plane. {The red arrows indicate the time of the snaphots shown on figures \ref{fig:eprime_snapshots_vert1} and \ref{fig:eprime_snapshots_vert2}.}}
\end{figure}
%
\begin{figure*}
\begin{tabular}{cc}
\parbox{0.5\textwidth}{\includegraphics[width=0.5\textwidth]{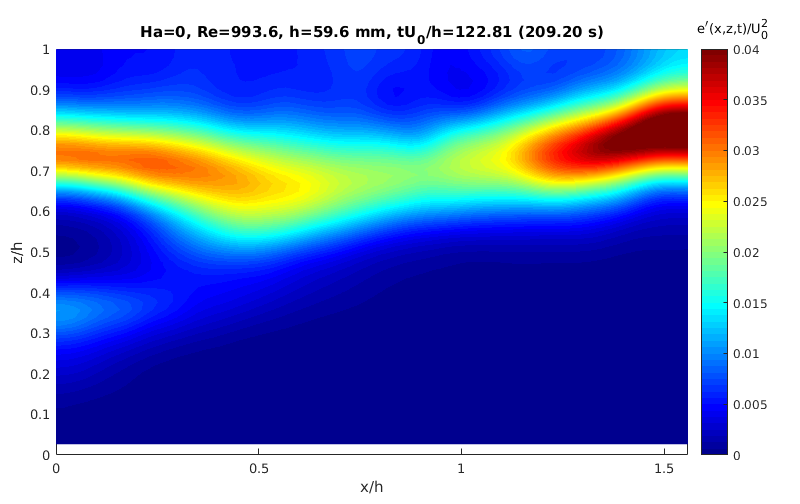}}&
\parbox{0.5\textwidth}{\includegraphics[width=0.5\textwidth]{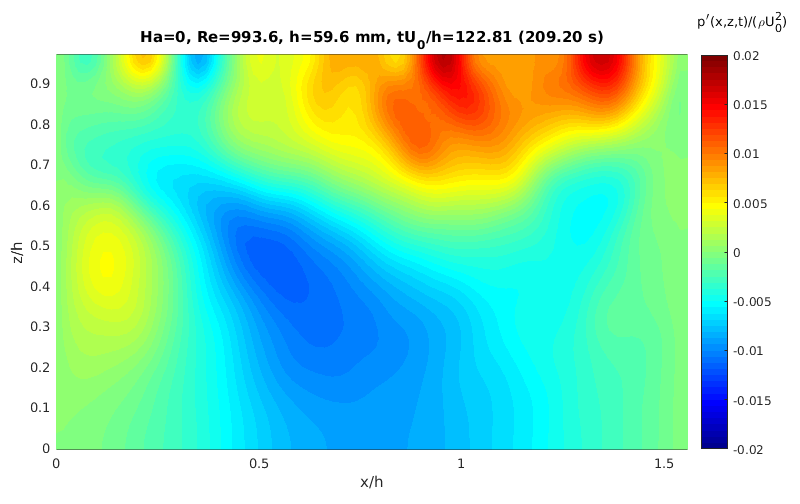}}\\
\parbox{0.5\textwidth}{\includegraphics[width=0.5\textwidth]{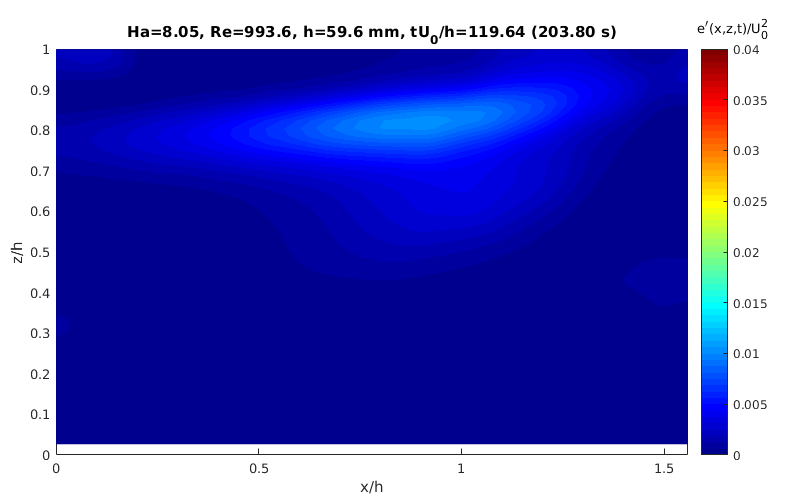}}&
\parbox{0.5\textwidth}{\includegraphics[width=0.5\textwidth]{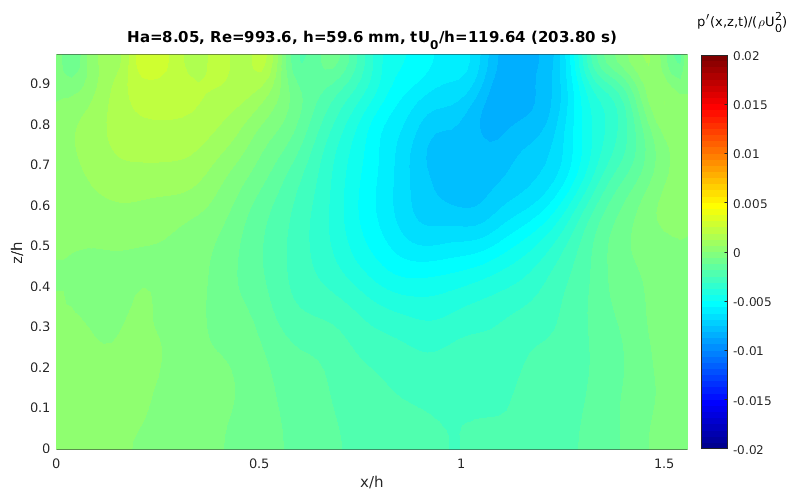}}\\
\end{tabular}
\caption{\label{fig:eprime_snapshots_vert1} Contours of local energy of the fluctuations (left) and pressure fluctuations (right) for $Re=993.6$, for $Ha=0$ and $Ha=8.05$ in the vertical plane. {The flow is directed from left to right.} The energy tends to be concentrated in localised perturbations of higher intensity, leading to extreme energy peaks during "rare events".}
\end{figure*}
\begin{figure*}
\begin{tabular}{cc}
\parbox{0.5\textwidth}{\includegraphics[width=0.5\textwidth]{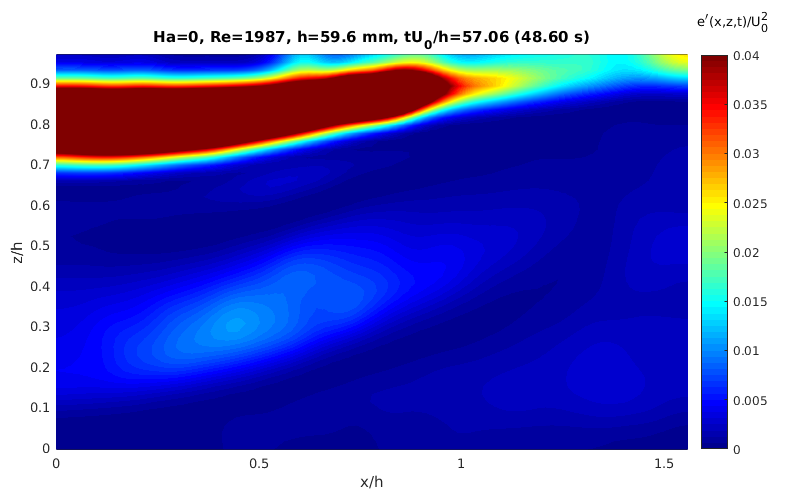}}&
\parbox{0.5\textwidth}{\includegraphics[width=0.5\textwidth]{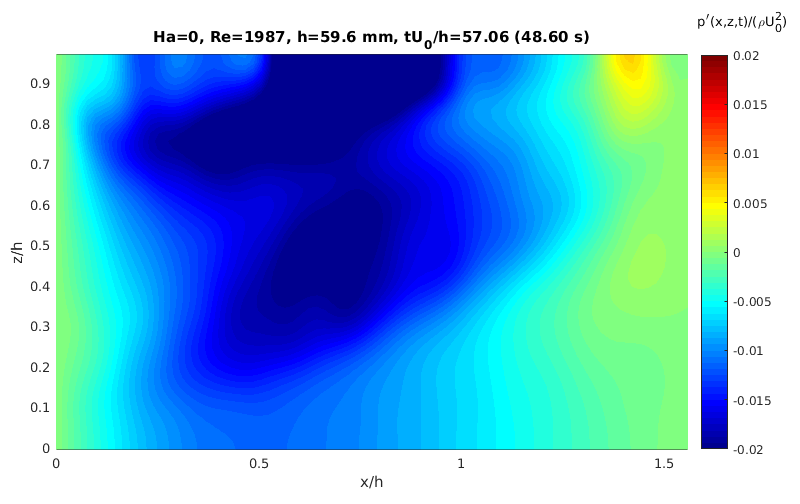}}\\
\parbox{0.5\textwidth}{\includegraphics[width=0.5\textwidth]{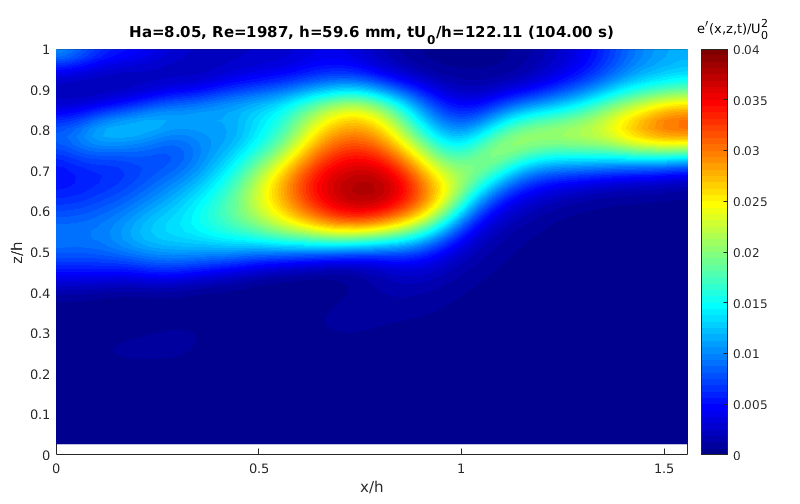}}&
\parbox{0.5\textwidth}{\includegraphics[width=0.5\textwidth]{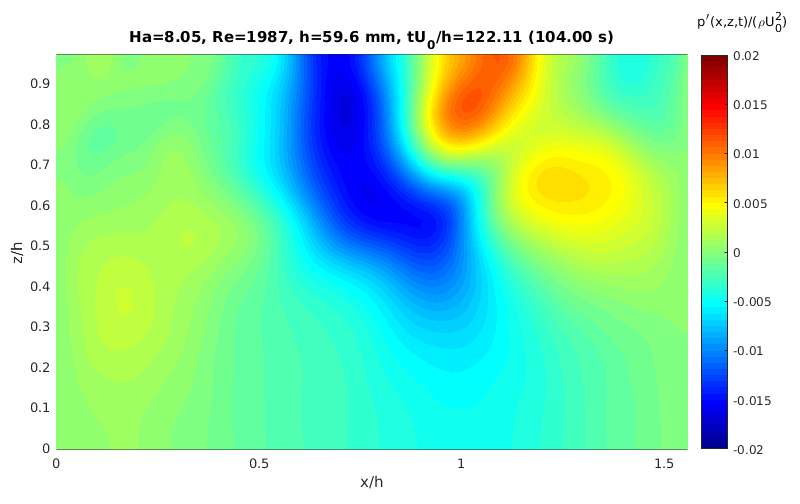}}
\end{tabular}
\caption{\label{fig:eprime_snapshots_vert2} Contours of local energy of the fluctuations (left) and pressure fluctuations (right) for $Re=1987$ (bottom), for $Ha=0$ and $Ha=8.05$ in the vertical plane. {The flow is directed from left to right.} The energy tends to be concentrated in localised perturbations of higher intensity, leading to extreme energy peaks during "rare events".}
\end{figure*}
%
%
Fig. \ref{fig:eprime_evol} shows records of $E(t)$ corresponding to the velocity profiles reported in sections 
\ref{sec:velocities}. Both the MHD and the non-MHD flows exhibit rare but extreme peaks of $E(t)$ 
reaching typically 10 times its average value {(at $t=48.52$ ($Ha=0$, $Re=1000$), $t=141.6$ ($Ha=8.05$, $Re=1000$), 
$t=57.53,152.9,157.2$ ($Ha=0$, $Re=2000$) and $t=35.22,68.8, 174.7$ ($Ha=8.05$, $Re=20000$)}. A closer inspection of the flow field during these peaks indicates that 
they are incurred by the passage of localised perturbations of much greater intensity than the background
fluctuations (visible on the snapshots  representing the contours of local energy $\mathbf u^{\prime 2}(x,z)$ on
figures \ref{fig:eprime_snapshots_vert1} \& \ref{fig:eprime_snapshots_vert2} (left)).
The full recordings used to
calculate the profiles of mean velocity and the \emph{rms} of its fluctuation, typically exhibit 2 or 3 of these events,
when a large number of them would be required to achieve statistical convergence. This would imply
extremely long time series (of the order of 10 hours).\\
Perturbations associated to extreme events also explain the two-maximum structure of the streamwise
perturbation profile, as movies and snapshots in Fig. \ref{fig:eprime_snapshots_vert1} \& \ref{fig:eprime_snapshots_vert2}  show that the most intense perturbations
navigate in the region where the plug flow in the bulk meets the high-shear region near the moving wall.
These perturbations are strongly anisotropic, typically elongated in a 1:10 ratio in the streamwise direction. Their 
upstream region comes very close to the moving wall, although the blind PIV region there hides
how close exactly. 
The upstream part of the perturbation dips into the bulk of the flow, giving the perturbations a wormy 
shape sitting at an acute angle with the streamwise direction.
The maximum in streamwise velocity fluctuations near $z/h\simeq0.2$ 
occurs because of similar but less intense perturbations located there.\\
The time evolution of $E(t)$ shows that its base value is reduced in the presence of magnetic field, for cases at the same value of $Re$. Movies indeed show that background fluctuations are damped across
the entire layer. {This is further quantified by calculating the time average of the energy of the fluctuations, excluding rare fluctuations whose energy exceeds the average calculated over all times by more than 100\%: at $Re=993.6$ (\emph{resp.} $Re=1987$), this quantity drops from 33.85 (\emph{resp.} 24.92) for $Ha=0$ to 
19.62 ( (\emph{resp.} 19.03 )for $Ha=8.05$.} At this
stage, it is however difficult to infer the impact of the magnetic field on the frequency and the
intensity of the rare, extreme events that skew the average {over all times}.\\ 
%
%
\begin{figure}
\centering
{\includegraphics[width=0.45\textwidth]{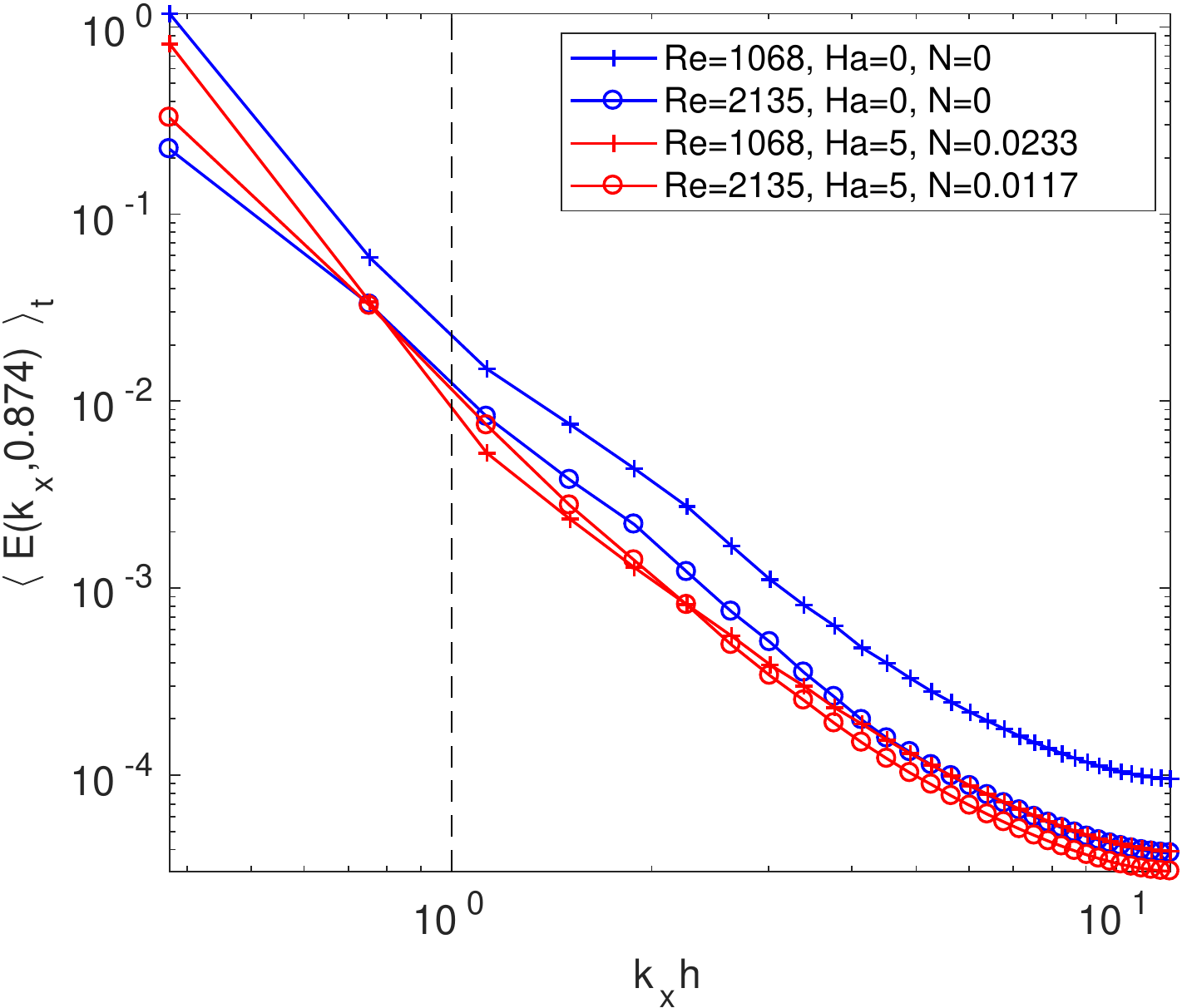}}\\
{\includegraphics[width=0.45\textwidth]{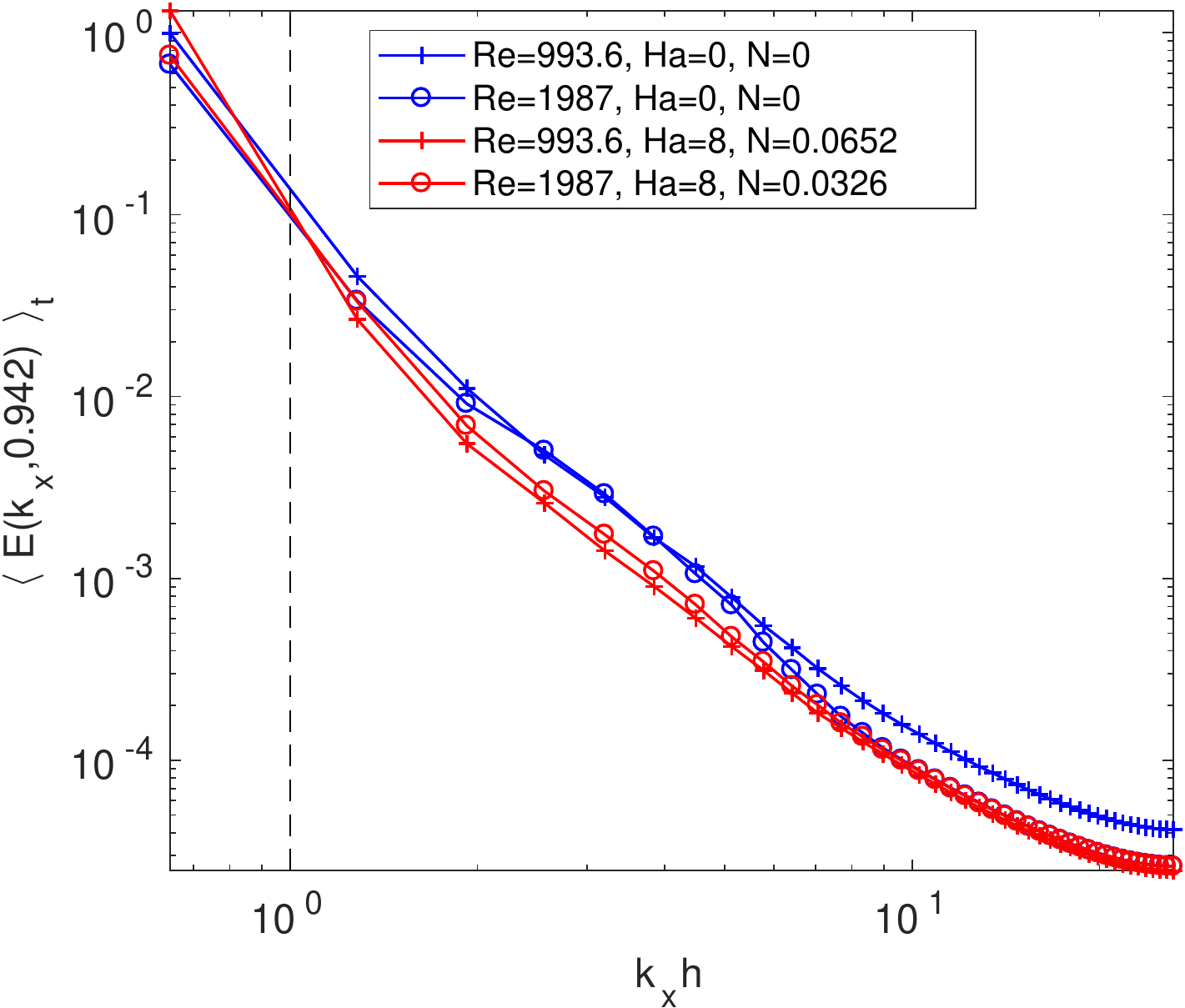}}\\
\caption{\label{fig:kx_spectra_vert} 1D energy spectra of energy fluctuations at $z=z_s$ for $Ha=0,5$ and 8.05 and several values of $Re$ with no perturbation grid at the inlet. The dashed line indicates $k_xh=1$ for which the streamwise size of the perturbation corresponds to the height of the channel.}
\end{figure}
Nevetherless, a better insight on the effect of the Lorentz force can be inferred from the spectral distribution of energy amongst streamwise wavenumbers $k_x$, since $x$ is the only invariant direction in the vertical plane. Also, since the analysis of velocity fields highlighted the importance of the region near the moving wall, we calculate spectra along a line
located at as close as possible to the moving wall, whilst still within the region of reliable PIV data, which is approximately 3 mm away from the wall. The corresponding non-dimensional coordinate $z_s/h$ therefore depends on the channel height ($z_s/h=0.873$, and 0.942 for $h=$32 mm and 59.6 mm respectively). Time averaged spectra $\langle E(z_s,k_x,t)\rangle_t$ are reported on Fig.
\ref{fig:kx_spectra_vert} for $Ha=$0, 5 and 8.05, for all values of $Re$ previously considered.\\
Spectra in all cases present a similar topology with a fairly constant slope of the order of $k_x^{-2}$ with a slight bump between $k_xh=2$ and $k_xh=3$.
Towards the high-end of the spectrum, the slope reduces. This is a known artefact of PIV measurements \cite{poelma2006_ef}.\\
The Lorentz force affects the spectra in two ways: although the general shape of the spectra is not significantly changed, the MHD spectra appear consistently a little steeper than their non-MHD counterpart. Furthermore, the energy is lower at all scales of the fluctuations, with the exception
of the very largest scale, which does not seem to follow a clear pattern when either $Ha$ and $Re$ are
varied.
Since the largest scales (of the streamwise size of the domain) cannot be resolved, this is not an
indication that larger scales (respectively $k_xh=$0.34, 0.37, 0.63 for $h=32$ mm, 35.5 mm and 59.6 mm) are not affected by the magnetic field. Furthermore, since $\langle E(z_s,k_x,t)\rangle_t$ decreases monotonically with
$k_x$, this implies that the main contribution to energy fluctuations would need a domain of larger
streamwise extension to be resolved. In a way, this is the spatial-spectral counterpart of the temporal
limitation observed on the statistical convergence of the \emph{rms} of velocity fluctuations. {Furthermore the structures undergo significant distortion through vortex-wall and vortex-vortex interaction during their transit through the visualisation windows. This makes it very difficult to overcome the 
limitation on window size with approaches based on Taylor's hypothesis.}\\
The analysis of  the spectra lifts part of the uncertainty raised during the analysis of the velocity profiles in
that they confirm a damping of velocity fluctuations at all scales by the magnetic field, that is, at first sight, commensurate with the interaction parameter $N$. 
\subsection{Pressure field}
\begin{figure}
\begin{tabular}{c}
{\includegraphics[width=0.5\textwidth]{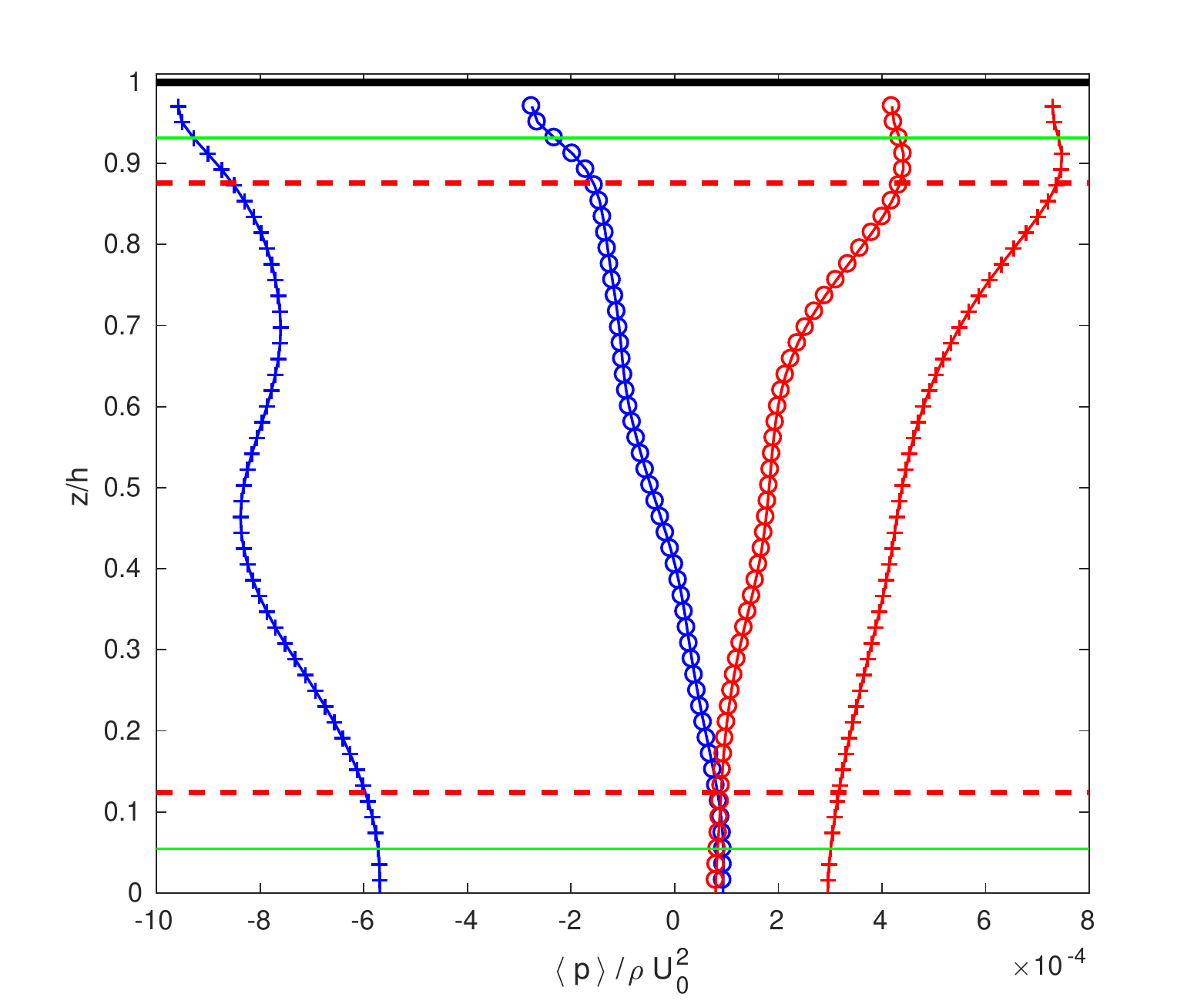}}\\
{\includegraphics[width=0.5\textwidth]{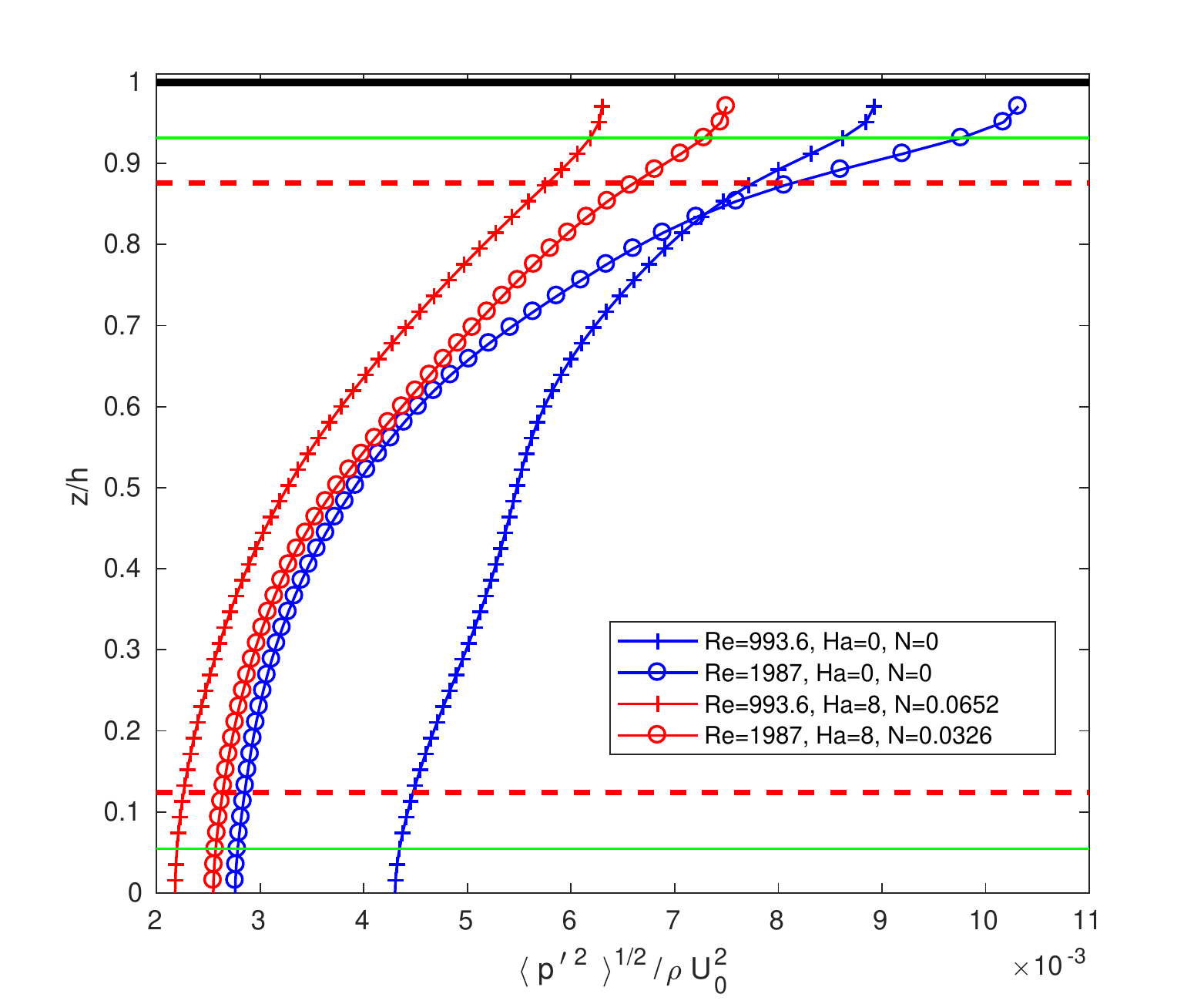}}
\end{tabular}
\caption{\label{fig:prms} Vertical profile of the average pressure (top) and \emph{rms} of 
pressure fluctuations (bottom) 
for $Ha=0$ and $Ha=8.05$, with no perturbation grid at the inlet. 
Dimensionally, Measurements were obtained with $h=59.6$ mm and $B=0$ T and $4$ T. 
{The red dashed line shows the location of the Hartmann layers, while the green solid
lines indicate the limits of reliability of the PIV data.}
}
\end{figure}
Fig. \ref{fig:prms} shows the vertical profiles of time-averaged pressure and time-\emph{rms} of its fluctuations for cases at $h=0.06$ m. 
In all cases, the average pressure is of the order of 10 times lower than the fluctuations (bearing in mind that the reference pressure at the inlet and outlet is 0), as expected for a flow without a driving pressure gradient. As such, the average value of the pressure is more an indication of the lack of statistical convergence due to the passage {of} large, rare perturbations. Furthermore, the pressure fluctuations are more important near the moving wall. 
This effect is also clearly visible on the snapshots of 
pressure fluctuations on figures \ref{fig:eprime_snapshots_vert1} \& \ref{fig:eprime_snapshots_vert2} (right): perturbations of local energy navigating in the region 
$z/h\simeq 0.8$ incur regions of excess pressure and deficit pressure around them. As a consequence, the maximum intensity of the 
pressure fluctuations tends to be located near the upper wall. This also explains why the average pressure has a higher absolute value in 
this region, as regions of excess or deficit pressure accumulate there over time but would require a large number of the large passing perturbations to cancel out on average.
 A second type of pressure perturbation occurs at the centre of high vorticity perturbations, this time under the form of a depression (visible in the example at $Ha=8.05$, $Re=1987$). These tend to be localised in the region $z\simeq 0.8$, and to a much lower extend $z\simeq 0.2$.\\ 
This phenomenology operates in the same way with and without magnetic field. As such the effect of the Lorentz force on the pressure field 
is indirect, through its action on the occurrence and intensity of the perturbations. 
\section{Conclusion}
This work produced several outcomes. First, we have reproduced a finite, plane MHD-Couette flow in a 
transparent and electrically conducting electrolyte. The device that achieved it operates in parameter range determined 
by the strength of the magnet used and the motor driving the belt that acts as the single moving wall.
With the equipment at hand, Hartmann and Reynolds numbers respectively up to approximately 10 and 12000 
can be reached. This makes it possible to investigate flow spanning regimes from laminar to turbulent.
For the sake of characterising the effect of the field in flows of intermediate complexity, we have 
focused on selected parameters in a transitional regime.\\
The main advantage of this setup is to provide access to 2D velocity and pressure fields in planes containing 
the streamwise direction and one spanwise direction. This was made possible by using electrolytes in high magnetic 
fields, but at the cost of several limitations: the low conductivity of the electrolyte (sulphuric acid) leaves 
the sort of high Hartmann numbers accessible in liquid metal experiments (such as $4\times10^4$ in \cite{bpddk2017_ef}) 
out of reach. The problem is exacerbated by the large size of the experiment that precludes fitting it in the bore of 
high field solenoidal magnets. Instead, pervading the entire domain with the magnetic field led us to place the rig 
in the weaker and less homogeneous stray field. Despite these limitations, we were able to implement the pressure PIV 
technique for the first time in a magnetohydrodynamic flow and measure the fluctuations of pressure induced by 
perturbations conveyed by the mean flow in the transitional regime of the MHD Couette flow. There is no doubt that 
future technological developments will help significantly mitigate the limitations we mentioned (for example guiding flux lines 
with appropriate polar pieces, could increase the effective field in the fluid approximately 3 fold, and even further with a split-pair 
magnet).\\
The analysis of the flow itself revealed that the Lorentz force creates a zone of strong shear in the vicinity of the moving tape of thickness scaling as $Ha^{-1}$. In a way, this is the first optical visualisation of a Hartmann layer, 45 years after Shercliff's elegant 
experimental evidence \cite{shercliff1965_jfm}. The {mean} flow in the bulk is, by contrast severely damped, as soon as the interaction parameter 
$N$  based on the tape velocity exceeds approximately a value as low as 0.02. Overall the Lorentz force reduces the flow rate at a given 
Reynolds number. {This value should however not be understood as a transition point but as a first estimate based on observations of the mean flow in the few cases presented here. A precise quantification of the effect of the Lorentz force would require an extensive parametric study.}\\
In the transitional regime, the flow was shown to be dominated by isolated perturbations navigating in two symmetric regions with 
respect to the mid-plane (at $z/h\simeq0.2$ and $z/h\simeq0.8$), with significantly more intense near-tape perturbations. This asymmetry  
in intensity may be attributed to the asymmetric inlet flow where the near-tape region undergoes a strong acceleration, which is absent 
near the bottom wall. These perturbations are strongly anisotropic, with significant elongation in the streamwise direction and seem 
attached to the tape. They incur regions of higher pressure impacting the upper moving wall only, as pressure fluctuations associated to the region $z\simeq0.2$ were low. Areas of low pressure also occur at the centre of perturbations with 
high vorticity in the upper part of the fluid layer but these are not as confined to the near-wall region as the regions of high pressure.  
The occurrence of these perturbations is rare at low $Re$ and becomes more frequent at higher Reynolds number (a fact that we have been able to verify on more measurements than presented here). As such, they play a significant role in the flow's transition to turbulence, and deserve a more systematic analysis. Their scarcity makes it very difficult to obtain converged statistics. 
Nevertheless, power density spectra do reveal that the Lorentz force 
indeed damped all scales we could detect in this setup and in this range of parameters, even with 
interaction parameters as low as  $N\simeq 10^{-2}$. 
{Indeed, both the cases shown here and other preliminary tests suggest that turbulent fluctuations may be affected at even lower values of the interaction parameter. The reason is most likely that their relative intensity remains low even at Reynolds numbers of up to at least $10^4$, so that interaction 
parameters built on them does not drop significantly below the order unity. This provides good hope that MHD turbulence of reasonable intensity may be accessible with this setup, especially if a stronger magnet can be used.}


\end{document}